\documentclass[aps,prb,twocolumn,groupedaddress,10pt]{revtex4-2}
\usepackage{graphicx}
\usepackage{dcolumn}
\usepackage{bm}
\usepackage[utf8]{inputenc}
\usepackage[T1]{fontenc}
\usepackage{mathptmx}
\usepackage{etoolbox}
\usepackage{float}
\usepackage{placeins}

\usepackage[utf8]{inputenc}
\usepackage[T1]{fontenc}
\usepackage{graphicx}
\usepackage{dcolumn}
\usepackage{bm}
\usepackage{newtxtext}
\usepackage{newtxmath}
\usepackage{multirow}
\usepackage{soul}
\usepackage[version=4]{mhchem}
\usepackage{hyperref}

\makeatletter
\def\@email#1#2{%  
  \endgroup
  \patchcmd{\titleblock@produce}
  {\frontmatter@RRAPformat}
  {\frontmatter@RRAPformat{\produce@RRAP{*#1\href{mailto:#2}{#2}}}\frontmatter@RRAPformat}
  {}{}
} % End of email definition
\makeatother

\begin{document}

\preprint{AIP/123-QED}

\title{Gate-tunable charge-spin interconversion in few-layer graphene/sputtered heavy metal heterostructures}

\author{Zhendong Chi$^{1,\dagger}$, Eoin Dolan$^{1,2,\dagger}$, Haozhe Yang$^{1}$, Beatriz Mart\'in-Garc\'ia$^{1,3}$, Marco Gobbi$^{3,4}$, Luis E. Hueso$^{1,3}$, and F\`elix Casanova$^{1,3,*}$}

\affiliation{$^1$CIC nanoGUNE BRTA, 20018 Donostia-San Sebastian, Basque Country, Spain}
\affiliation{$^2$Departamento de Polímeros y Materiales Avanzados: Física, Química y Tecnología, University of the Basque Country (UPV/EHU), 20018 Donostia-San Sebastian, Basque Country, Spain}
\affiliation{$^3$IKERBASQUE, Basque Foundation for Science, 48009 Bilbao, Basque Country, Spain}
\affiliation{$^4$Centro de F\'isica de Materiales (CSIC-EHU/UPV) and Materials Physics Center (MPC), 20018 Donostia-San Sebastian, Basque Country, Spain}

\affiliation{$\dagger$ These authors equally contributed to this work}

\email[Corresponding author ]{f.casanova@nanogune.eu}

\date{\today}

\begin{abstract}
Spintronics has emerged as a promising field for next-generation devices, offering functionalities beyond complementary metal-oxide-semiconductor (CMOS). A critical challenge in spintronics is to develop systems that can efficiently generate spin currents and enable their long-distance transport. Here, we demonstrate a graphene (Gr)/heavy metal (HM) heterostructure system that combines strong charge-spin interconversion efficiency, induced by the spin Hall effect, with a long spin diffusion length. By employing an industry-friendly magnetron sputtering technique, we deposit HM layers onto few-layer Gr while minimizing structural damage. The proximity effect from the HM enhances the spin Hall angle of Gr while limiting the reduction in its spin diffusion length. Additionally, the spin Hall angle can be tuned via an applied gate voltage, offering high controllability of the system. Importantly, these properties are observed across heterostructures composed of different HMs, indicating the generality of this approach. Our findings establish Gr/HM heterostructures as a scalable and versatile platform for spin current generation, paving the way for advanced spintronic devices with high efficiency, long spin propagation, and straightforward fabrication processes.
\end{abstract}

\maketitle

\thanks{*These authors equally contributed to this work.}

\section{Introduction}

The efficient generation and manipulation of spin currents lies at the heart of spintronics, a field whose potential applications extend far beyond conventional electronics~\cite{02Garello2018-kv, 05Incorvia2024-uv, 06Fert2024-gv}. Charge-spin interconversion (CSI) is particularly significant because it enables the creation of spin currents from charge currents, thus simplifying the integration of spintronics with traditional electronic systems~\cite{03Manipatruni2019-xy, 04Pham2020-nl, 07Kimura2007-em, 08Manchon2015-uv, 09Miron2011-gl, 10Liu2012-fm}. Among the various CSI phenomena, the spin Hall effect (SHE) has been widely investigated and is typically observed in materials with strong spin-orbit coupling (SOC), such as 5$d$ heavy metals (HMs)~\cite{01Sinova2015-ip, 07Kimura2007-em, 12Kim2013-sn, 13Chi2020-mv}. For large-scale integration, not only the magnitude of the SHE, characterized by the spin Hall angle ($\theta_\mathrm{SH}$) but also the spin diffusion length ($\lambda_\mathrm{s}$)---the effective distance over which spin current can propagate---is essential~\cite{14Liu2011-dz, 15Isasa2015-it, 16Sagasta2016-pt, 17Sagasta2017-ms, 18Sagasta2018-xd, Morota2011-hl, 11Rojas-Sanchez2014-wz, 49Kim2016-jm}. While HMs exhibit a strong SHE, their substantial SOC strongly reduces $\lambda_\mathrm{s}$ by causing significant spin relaxation. 

Graphene (Gr), the most recognized two-dimensional (2D) material, has recently attracted substantial attention in spintronics due to its extraordinarily long spin diffusion length---as high as tens of micrometers~\cite{19Geim2013-ff, 20Han2014-ut, 21Avsar2020-tf, 22Sierra2021-yi, 23Ingla-Aynes2015-or, 24Ingla-Aynes2016-gx, 25Yan2016-qj, Drogeler2016-eq, Bisswanger2022-bu}. Moreover, as a 2D material, its properties can be easily tuned by an external gate voltage, highlighting its promise for both electronic and spintronic applications~\cite{26Guimaraes2014-rm, 27Leutenantsmeyer2018-bq, 28Ingla-Aynes2021-dv}. However, Gr's intrinsic SOC is remarkably weak, as it consists solely of light carbon atoms, making its inherent SHE negligible.

To capitalize on Gr's long spin diffusion length for spintronic applications, considerable effort has been devoted to enhancing its SHE by adding adlayers containing HM elements on the Gr surface. Due to proximity effects, the strong SOC of an adjacent layer can imprint onto Gr, thus inducing CSI ~\cite{29Garcia2017-dx, 30Garcia2018-pl, 31Offidani2017-ep}. These improvements have been observed in a variety of systems, including Gr/transition metal dichalcogenide (TMD) heterostructures fabricated by mechanical transfer~\cite{37Ghiasi2019-ls, 38Safeer2019-cd, Yang2024-kv, 32Yang2024-sa, 34Safeer2019-bl,35Benitez2020-mn, 36Herling2020-xi, 39Zhao2020-jx, 33Chi2024-je, Li2020-yh, Khokhriakov2020-qq, Hoque2020-ty, Hoque2021-dv, 44Camosi2022-bu, 45Ingla-Aynes2022-lw, 46Ontoso2023-hj} and Gr/HM heterostructures where the HM is deposited via evaporation~\cite{40Safeer2020-io, 41Yang2023-ji}. Although these systems demonstrate CSI, the spin diffusion length often decreases significantly~\cite{35Benitez2020-mn}, making it challenging to combine both a robust CSI efficiency and a long spin diffusion length for practical devices. In particular, for magnetoelectric spin–orbit (MESO) logic devices~\cite{03Manipatruni2019-xy, 04Pham2020-nl}, the key figure of merit is the product, $\lambda_\mathrm{CSI}=\theta_\mathrm{SH} \times \lambda_\mathrm{s}$, and therefore achieving proximitized graphene with a long $\lambda_\mathrm{s}$ is highly desirable.

In this paper, we report the successful realization of a strong SHE while retaining a long spin diffusion length in Gr/HM heterostructures, achieved via an industry-friendly magnetron sputtering process. By optimizing the Gr thickness and using a grazing-incidence sputtering technique, we minimize structural damage to Gr during deposition. Our results indicate that the HM layer substantially increases the $\theta_\mathrm{SH}$ of Gr, while causing only a small reduction in its spin diffusion length. Furthermore, we demonstrate that the $\theta_\mathrm{SH}$ can be effectively tuned by applying a gate voltage, enabling highly controllable CSI efficiency. This efficiency is on par with that observed in Gr/TMD heterostructures. Significantly, this phenomenon is consistently observed for various HMs, underscoring its broad applicability. Our findings thus offer a promising strategy for the generation of spin currents, paving the way for industrially feasible, cutting-edge spintronic technologies.

\section{Device Fabrication and Characterization}
Gr flakes of varying thicknesses were exfoliated onto highly \(n\)-type Si substrates with a 300-nm-thick SiO\(_2\) layer. The number of Gr layers was determined using contrast imaging via an optical microscope and confirmed through Raman spectroscopy. Gr/HM Hall bar devices were fabricated in multiple steps.

The Hall bar structure was defined on the Gr using e-beam lithography. A 20-nm-thick Al layer was deposited as a hard mask to protect the Hall bar Gr, while the remaining area was etched using Ar-O\(_2\) reactive ion etching. The Al mask was subsequently removed using MF319. The Gr Hall bar was annealed at \(400^\circ\)C for 1 hour to eliminate residual polymers on the surface. A second e-beam lithography step defined the centre square area at the Hall bar intersection, and ultrathin HM layers (Ta, W, and Pt) were deposited via magnetron sputtering with a grazing angle (less than 10$^\circ$ from the surface) in the case of Ta and Pt, and direct angle but low power in the case of W, in order to minimize plasma damage on the Gr. The specific parameters were an Ar pressure of 3 Pa for all depositions, with a DC power of 250 W for Ta, 80 W for Pt, and 10 W for W. 

The HM layer thicknesses were 1 nm, 1 nm, and 0.5 nm for Ta, W, and Pt, respectively. The 1-nm-thick Ta and W layers were fully oxidized, while the 0.5-nm-thick Pt film remained discontinuous. The complete oxidation of the Ta and W is intentional, as it ensures that the spin current is confined to the graphene channel, which prevents spin absorption and additional conversion signals that would arise in a thicker, conductive metal film. In the case of Pt, oxidation is not possible, and the metal remains conductive, however since the layer is discontinuous the spintransport should remain dominated by the graphene.

Non-magnetic (NM) Ti (5 nm)/Au (35 nm) electrodes were fabricated to electrically contact the Gr using standard e-beam lithography, evaporation, and lift-off processes. To inject and detect spin currents, TiO\(_x\)/Co/Au ferromagnetic (FM) electrodes were deposited. First, a 0.3-nm-thick Ti layer was evaporated and oxidized in ambient conditions to form TiO\(_x\), followed by deposition of 35-nm-thick Co and 15-nm-thick Au layers. The Au cap prevents Co oxidation. The FM electrodes had nominal widths of either 150 or 350 nm to induce different magnetic shape anisotropies.

Raman spectroscopy was performed using an Alpha 300R Confocal Raman WITec microscope with a 532-nm laser as the excitation source, a 600 1/mm diffraction grating, and a 100× objective lens. Pristine Gr and Gr/HM heterostructures without microfabrication were characterized to validate the layer quality.

Electrical measurements were conducted in a physical property measurement system (PPMS) by Quantum Design. Measurements were performed at various temperatures and magnetic fields applied along different directions. A DC reversal technique, employing a Keithley 2182 nanovoltmeter and a 6221 current source, was used to measure resistance. A back-gate voltage ($V_\mathrm{G}$) was applied using a Keithley 2636B source meter to control the Fermi level of the Gr.

\section{Structural and electrical characterization}

\begin{figure}[h!]
    \centering
    \includegraphics[width=0.48\textwidth]{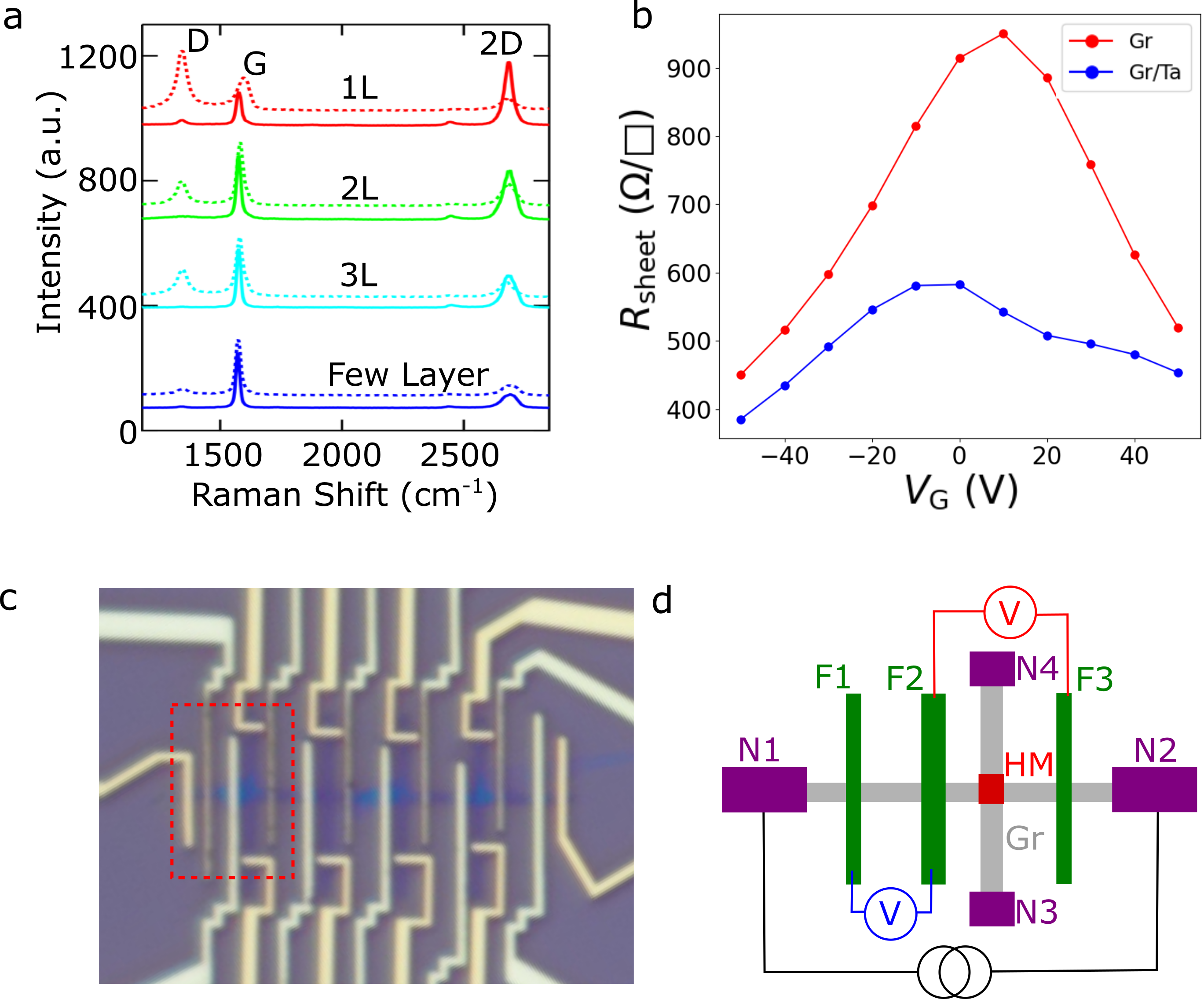} % Adjust figure path as needed
    \caption{\label{fig:device_schematic}
        (a) Raman spectra of Gr before (solid lines) and after (dashed lines) 1-nm-thick Ta sputtering deposition for monolayer (1L, red), bilayer (2L, green), trilayer (3L, cyan), and few-layer (blue) Gr. The spectra are normalized and vertically shifted for clarity. The characteristic peaks (D, G, and 2D) are labeled. (b) \( V_{\text{G}} \)-dependent sheet resistance (\( R_{\text{sheet}} \)) of pristine Gr (blue) and Gr/Ta (red) measured at 50~K. (c) Optical image of Gr/Ta heterostructure devices for transport measurements. (d) Schematic of the section of the device measured, corresponding to the outlined section in (c), showing labelled FM and NM contacts, the pristine Gr and the centre of the cross proximitized with the HM. The contacts corresponding to the data in (b) are shown as the red and blue voltage measurements, respectively.
    }
\end{figure}

We first investigate the impact of sputtering deposition of HM elements on the quality of Gr. As a representative example, we employ a 1-nm-thick Ta layer to study the effect of the HM deposition process. Fig.~\ref{fig:device_schematic}a illustrates the Raman spectra of Gr with varying layer numbers. The solid and dashed curves correspond to spectra obtained before and after the deposition of the Ta layer, respectively. 

We observe that the intensity of the 2D peak decreases significantly, while the D peak appears following the sputtering deposition of the Ta layer, particularly when the Gr layer number is fewer than three. These results strongly suggest that atomic defects and disorder in Gr are enhanced due to the HM sputtering process.~\cite{42Malard2009-zn} Nevertheless, the change in Raman spectra before and after Ta sputtering deposition becomes negligible for few-layer Gr, where the layer number exceeds six, as confirmed by atomic force microscopy (AFM). This strongly suggests that the sputtering-induced damage is confined to the first few layers of Gr. Assuming that the top layers of the few-layer Gr are damaged, we maintain several layers of undamaged Gr underneath which facilitates spin transport. Thus, our investigations focus on devices featuring few-layer Gr, which are robust enough to maintain good spin transport properties after sputtering. Given the short range of the proximity effect in Gr, the proximity could be confined to the same layers which are damaged by sputtering, although the exact nature of the Gr/HM interface is very difficult to determine.

A typical Gr/Ta heterostructure device designed for local and non-local transport measurements is shown in Fig.~\ref{fig:device_schematic}c. The Gr is etched into Hall-bar crosses, and the Ta layer is deposited in the central part of these crosses. In subsequent sections, we will focus on the characterization of the heterostructure, circled in Fig.~\ref{fig:device_schematic}c. The FM and NM electrodes are denoted as F1--F3 and N1--N4 (Fig. 1d), respectively, and are used for injecting and detecting both spin and electrical currents.

To explore the impact of Ta layer deposition on the charge neutrality point (CNP) of Gr, we measured the Gr sheet resistance (\( R_{\text{sheet}} \)) as a function of $V_\mathrm{G}$ at 50~K, as shown in Fig.~\ref{fig:device_schematic}b. The measurement geometries are depicted in Fig.~\ref{fig:device_schematic}d. The blue and red curves represent the \( R_{\text{sheet}} \) of pristine Gr and the Gr/Ta regions, respectively. We used 3D FEM simulation to isolate the contribution of the Gr/Ta region (See Note S3 \cite{supinfo}) since the experimental measurement contains contributions from both the Gr and Gr/Ta (Fig.~\ref{fig:device_schematic}d). It is evident that, following the Ta sputtering deposition, the Gr/Ta region exhibits a decrease in \( R_{\text{sheet}} \) along with a shift of the CNP from $V_\mathrm{G}=+10\,V$ in the pristine Gr to $V_\mathrm{G}=-5\,V$ in the Gr/Ta region. These changes are attributed to carrier doping induced by the deposition of the Ta layer onto the Gr.

\section{Spin Transport  Characterization}
\begin{figure}[h!]
    \centering
    \includegraphics[width=0.48\textwidth]{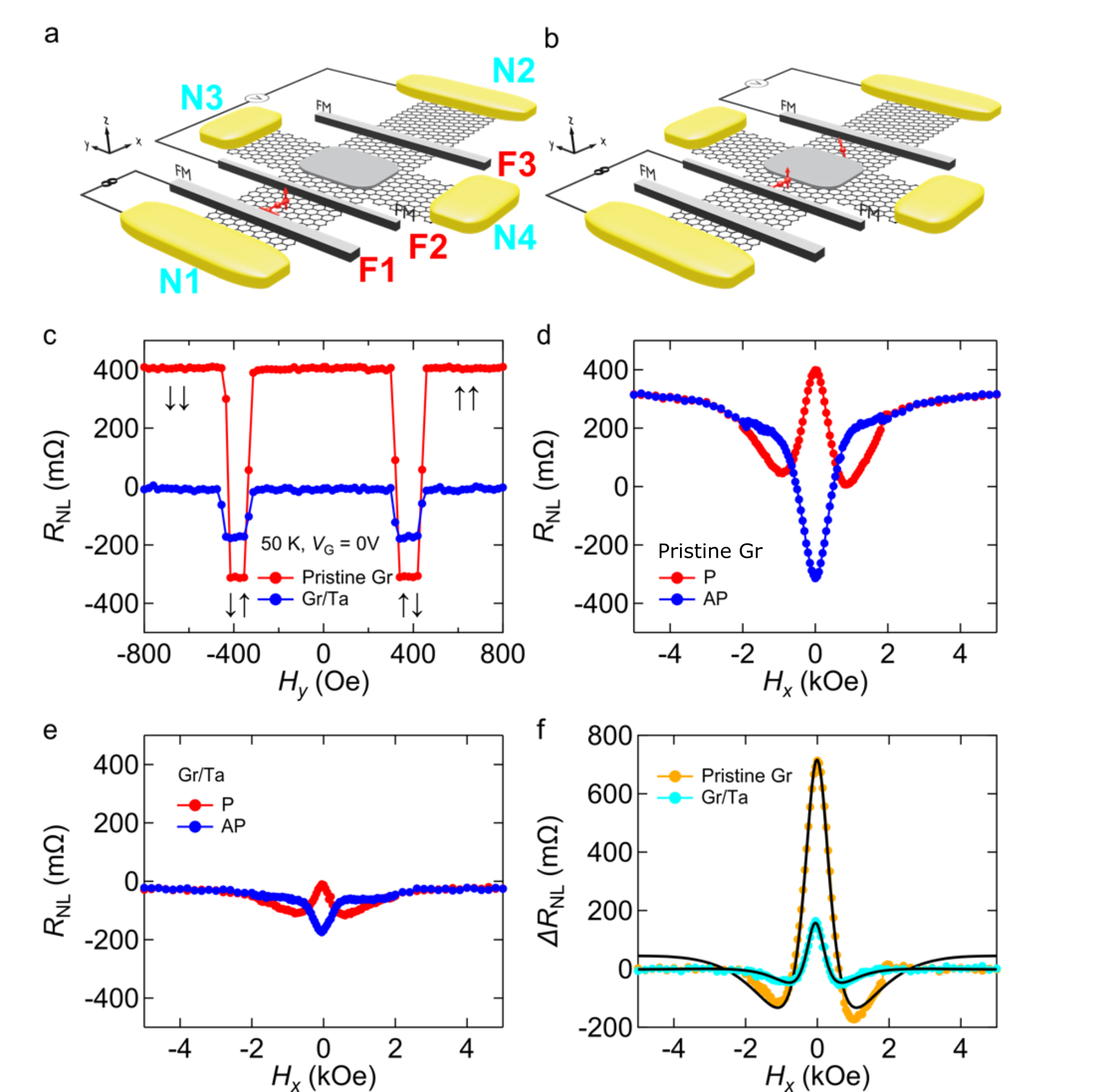}% Adjust figure path as needed
    \caption{\label{fig:LSV}
        (a, b) Schematic of the device with non-local spin transport measurement geometry for pristine Gr (a) and Gr/Ta heterostructure (b). 
        (c) Non-local resistance $R_\mathrm{NL}$  of the LSV measured as a function of $H_\mathrm{y}$. The red and blue lines correspond to results for pristine Gr and Gr/Ta heterostructure, respectively. 
        (d, e) Hanle precession measurements of pristine Gr (d) and Gr/Ta heterostructure (e) measured as a function of $H_\mathrm{x}$. The red and blue curves represent measurements with the two FM electrodes of the LSV in parallel ($R_\mathrm{NL}^\mathrm{P}$) and antiparallel ($R_\mathrm{NL}^\mathrm{AP}$) configurations, respectively. 
        (f) The difference in $R_\mathrm{NL}$  values (\( \Delta R_{\text{NL}} = R_{\text{NL}}^\mathrm{P} - R_{\text{NL}}^\mathrm{AP} \)) extracted from (d) (orange) and (e) (cyan). The black lines represent the fitting using a 3D FEM spin diffusion model. All data were obtained at 50~K without applied gate voltage.
    }
\end{figure}

After confirming the structural and electrical properties of the Gr/Ta heterostructures, we now present the spin transport characteristics of our samples. To investigate these properties, we employed non-local magnetotransport measurements using a DC reversal technique. The measurement configurations are shown in Figs.~\ref{fig:LSV}a and~\ref{fig:LSV}b, corresponding to lateral spin valves (LSVs) for pristine Gr and Gr/Ta heterostructures, respectively.

In Fig.~\ref{fig:LSV}a, a DC current (\( I_{\text{dc}} \)) is injected between electrodes F1 and N1, introducing a spin-polarized current from the FM electrode F1 into the Gr layer. This creates a pure spin current diffusing toward electrode F2 due to the high spin chemical potential at F1. The spin current is subsequently detected by another FM electrode, F2, and the resulting non-local voltage (\( V_{\text{NL}} \)) is measured between F2 and N3 using a nanovoltmeter. The corresponding non-local resistance is defined as $R_{\text{NL}} \equiv {V_{\text{NL}}}/{I_{\text{dc}}}$.

In this configuration, spin current transport occurs exclusively within the Gr layer, enabling characterization of the spin transport properties of pristine Gr. In Fig.~\ref{fig:LSV}b, the injection and detection FM electrodes are shifted to F2 and F3, respectively. In this case, the spin current traverses both the Gr and the Gr/Ta heterostructure. This configuration contains contributions from both the Gr/Ta region and the Gr region. The experimental data therefore depends on the effective value of the combined region, but we can isolate the behaviour in the pure Gr/Ta region using 3D finite element method (FEM) simulations (see Note S7 \cite{supinfo}).

Non-local spin valve measurements are performed by detecting $R_\mathrm{NL}$  while sweeping a magnetic field along the \( y \)-axis ($H_\mathrm{y}$).\cite{43Yan2017-fw} Since the easy axes of the FM electrodes are aligned along the \( y \)-axis and their shape anisotropies differ, sweeping $H_\mathrm{y}$ enables magnetization switching of the FM electrodes at relatively small field strengths, with each electrode responding at different field magnitudes.

The $R_\mathrm{NL}$ for both pristine Gr and the Gr/Ta heterostructure, measured at 50~K and no gate voltage, are plotted in Fig.~\ref{fig:LSV}c as a function of $H_\mathrm{y}$. A clear distinction is observed between the parallel (P) and anti-parallel (AP) magnetization configurations of the FM electrodes, confirming successful spin injection into the Gr layer, and well-behaved switching of the FMs. However, compared to the pristine Gr, the difference in $R_\mathrm{NL}$  between the P and AP configurations is significantly smaller in the Gr/Ta heterostructure. This difference corresponds exactly to the difference in magnitude of the Hanle precession measurements (Figs. \ref{fig:LSV}d, \ref{fig:LSV}e) in each region at $H_\mathrm{x}=0$. Since the distance between the electrodes is the same in each case, this difference is a combination of the effective $\lambda_\mathrm{s}$ in each region, and the spin injection efficiency of the different FM electrodes used in each measurement. This efficiency is characterized by the interface polarization, $P_\mathrm{i}$, which differs by a factor of approximately two between the two cases (see Note S6 \cite{supinfo}). Since the measured $R_\mathrm{NL}$ depends on $P_\mathrm{i}^2$, this difference in $P_\mathrm{i}$ almost entirely explains the difference in signal, without considering absorption or scattering due to the HM affecting $\lambda_\mathrm{s}$ (see Note S6 \cite{supinfo}).

To extract the spin transport parameters, \( D_\mathrm{s}, \tau_\mathrm{s}, \) and \( \lambda_\mathrm{s} \), where \( D_\mathrm{s} \) is the spin diffusion coefficient, \( \tau_\mathrm{s} \) is the spin relaxation time, and \( \lambda_\mathrm{s} \) is the spin diffusion length (defined by \( \lambda_\mathrm{s} = \sqrt{D_\mathrm{s} \tau_\mathrm{s}} \)), as well as $P_\mathrm{i}$, we perform the Hanle precession measurements using the configurations in Figs.~\ref{fig:LSV}a and \ref{fig:LSV}b, with applied field $H_\mathrm{x}$. Since the easy axes of the FM electrodes are along the \( y \)-axis, the initial magnetization of the injected spin current also lies along the \( y \)-axis. When $H_\mathrm{x}$ is applied, i.e., along the hard axes of the FM electrodes, the spin current undergoes precession in the \( y \)-\( z \) plane during diffusion through the Gr layer. Depending on the initial magnetization alignment of the FM electrodes (P or AP), the measured $R_\mathrm{NL}$  displays different trends. By analysing the difference the spin transport parameters can be extracted.

Figures~\ref{fig:LSV}d and \ref{fig:LSV}e present Hanle precession results measured at 50~K and no applied gate voltage for pristine Gr and the Gr/Ta heterostructure, respectively, showing $R_\mathrm{NL}$  as a function of $H_\mathrm{x}$. The red and blue curves correspond to measurements with initial magnetization alignments in P ($R_\mathrm{NL}^\mathrm{P}$) and AP ($R_\mathrm{NL}^\mathrm{AP}$) configurations, respectively. As $H_\mathrm{x}$ increases, the two curves converge due to the saturation of the FM electrode magnetizations along the \( x \)-axis, as well as spin dephasing due to precession in the external magnetic field. Figure~\ref{fig:LSV}f illustrates the difference between $R_\mathrm{NL}^\mathrm{P}$ and $R_\mathrm{NL}^\mathrm{AP}$, defined as \( \Delta R_{\text{NL}} \). A clear difference of magnitude is visible between the two curves, but as mentioned previously this can be explained by interface properties, rather than changing spin transport characteristics. To extract and compare spin transport parameters we fit the two datasets using a 3D FEM spin diffusion model, where we fit the shape of the curve using $D_\mathrm{s}$ and $\tau_\mathrm{s}$, followed by fitting the magnitude using $P_\mathrm{i}$ (see Note S5 and S6 \cite{supinfo}). Although we perform the fitting using $D_\mathrm{s}$ and $\tau_\mathrm{s}$, we focus on $\lambda_\mathrm{s}$ here since it is the more pertinent quantity in the calculation of CSI efficiency. The fits using this method are shown as the black curves in Fig. \ref{fig:LSV}f.

\begin{figure}[h!]
    \centering
    \includegraphics[width=0.48\textwidth]{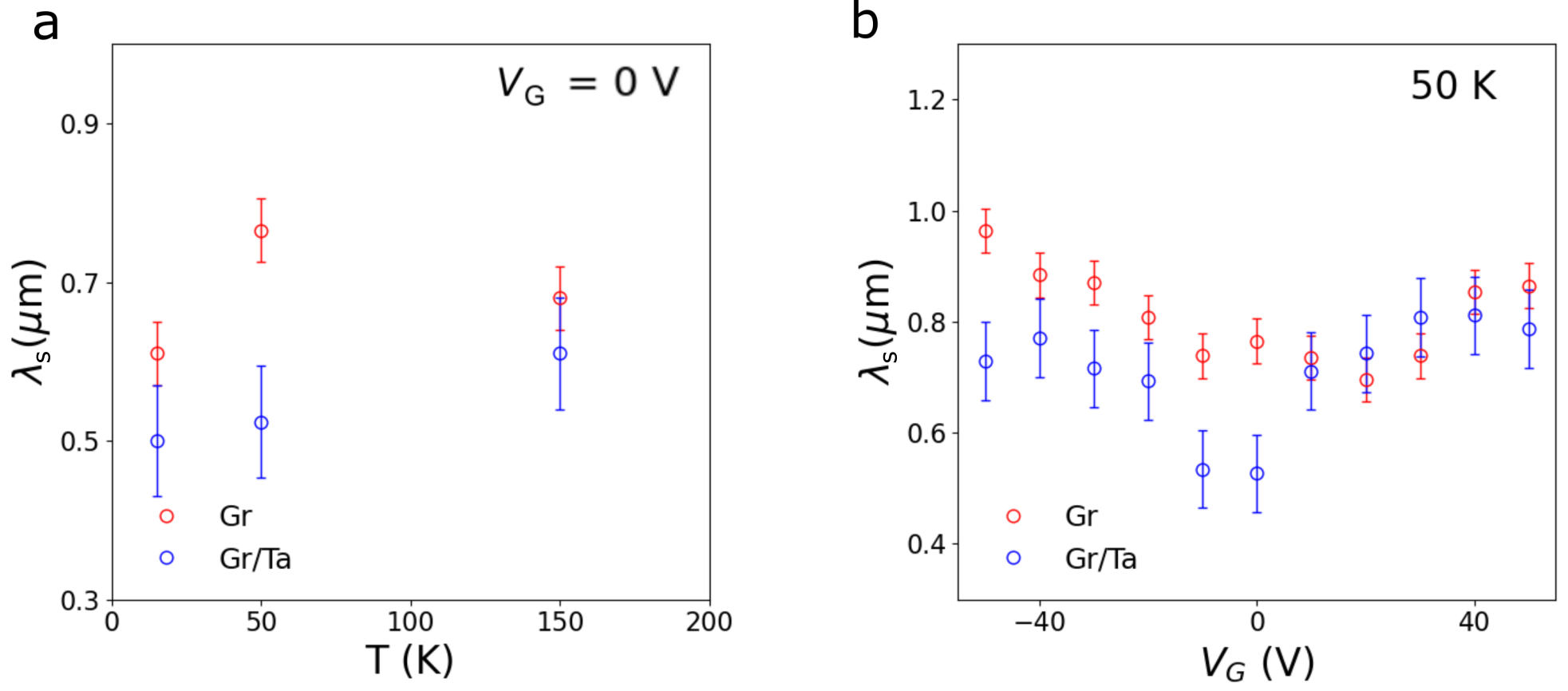}% Adjust figure path as needed
    \caption{\label{fig:Summary}
        Temperature (a) and gate voltage (b) dependences of spin diffusion length ($\lambda_\mathrm{s}$) for pristine Gr (red) and the Gr/Ta heterostructure (blue). The values in (a) are extracted from measurements conducted without an applied gate voltage, while the values in (b) are determined at 50~K.
    }
\end{figure}

The temperature and $V_\mathrm{G}$ dependence of $\lambda_\mathrm{s}$ for pristine Gr and the Gr/Ta heterostructure are summarized in Figs.~\ref{fig:Summary}a and~\ref{fig:Summary}b, respectively. There is no clear trend in the temperature dependence, however when tuning $V_\mathrm{G}$ at a constant temperature of 50~K, $\lambda_\mathrm{s}$ in both regions exhibits a minimum near the CNP of each region and increases with increasing $V_\mathrm{G}$. This behaviour matches well with the trend of $R_\mathrm{sheet}$ shown in Fig.~\ref{fig:device_schematic}b and aligns with observations in other Gr/TMD heterostructure systems.\cite{33Chi2024-je, 41Yang2023-ji} Notably, $\lambda_\mathrm{s}$  in pristine Gr is consistently, although not significantly, larger than in the Gr/Ta heterostructure across most conditions. This reduction is attributed to spin relaxation in the heterostructure, a phenomenon also reported in various Gr/TMD heterostructures.~\cite{35Benitez2020-mn, 34Safeer2019-bl, 40Safeer2020-io}

\section{Charge-Spin Interconversion}

\begin{figure}[h!]
    \centering
    \includegraphics[width=0.48\textwidth]{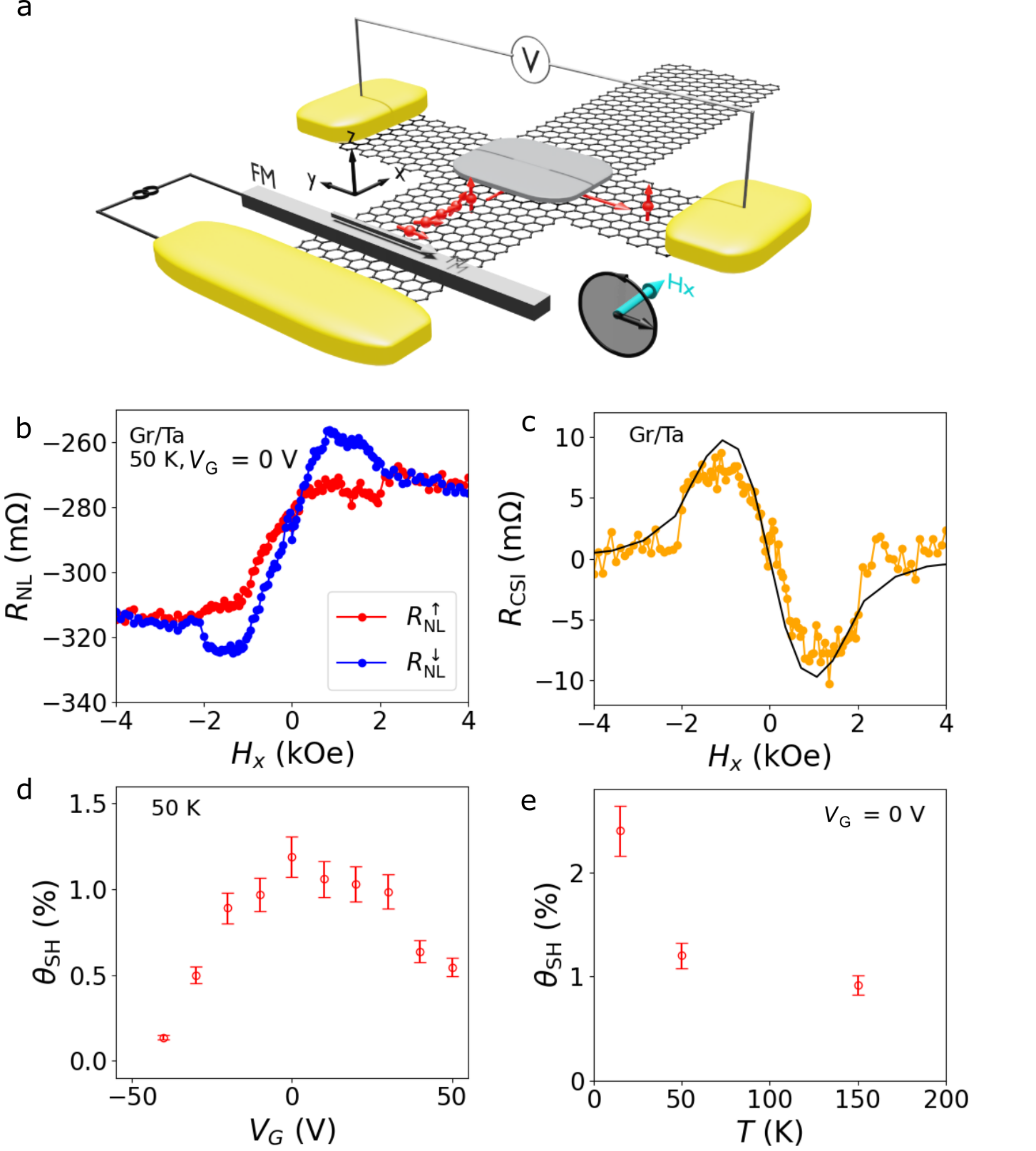}% Adjust figure path as needed
    \caption{\label{fig:CSI}
        (a) Schematic illustration of the CSI measurement geometry. 
        (b) Non-local resistance $R_\mathrm{NL}$  for CSI measured as a function of $H_\mathrm{x}$ in the Gr/Ta heterostructure at 50~K without external gate voltage. The red (\( R_{\text{NL}}^\uparrow \)) and blue (\( R_{\text{NL}}^\downarrow \)) curves represent measurements with the initial magnetization of the FM electrodes along the \( +y \) and \( -y \) directions, respectively. 
        (c) The CSI resistance \( R_{\text{CSI}} \) (yellow) extracted from the difference of the red and blue curves in (b). The black curve represents the fitting results. 
        (d, e) Gate voltage (d) and temperature (e) dependent spin Hall angle (\( \theta_{\text{SH}} \)) in the Gr/Ta heterostructure.
    }
\end{figure}

After confirming the spin transport properties of the Gr/Ta heterostructure, we proceed to investigate its CSI properties. The non-local CSI measurement geometry is illustrated in Fig.~\ref{fig:CSI}a. In this measurement, a spin-polarized current is injected from the FM electrode (F2 in the current system) using a DC current ($I_{\text{dc}}$), leading to the diffusion of a pure spin current towards the heterostructure. Due to the orthogonal relationship between spin and charge currents in the CSI process, if CSI occurs in the Gr/Ta heterostructure, a charge current is generated along the $y$-axis. This charge current is detected using a nanovoltmeter connected to two NM electrodes (N3 and N4) located on opposite sides of the Gr/Ta heterostructure. The measured voltage ($V_{\text{NL}}$), is again normalized to the applied current, $R_{\text{NL}} \equiv {V_{\text{NL}}}/{I_{\text{dc}}}$.

During the measurement, a magnetic field ($H_x$) is applied to induce precession of the spin current within the $y$-$z$ plane. As $H_x$ increases, the magnetization of the FM electrode gradually aligns along the $x$-axis, creating a projection of the spin current polarization onto the $x$-axis. This configuration allows probing the components of spin current polarization along the $x$-, $y$-, and $z$-axes.

Fig.~\ref{fig:CSI}b illustrates the CSI measurement results for the Gr/Ta heterostructure at 50~K without applied gate voltage. $R_\mathrm{NL}$  is plotted as a function of $H_\mathrm{x}$. The red and blue curves correspond to measurements taken with FM electrodes having opposite initial magnetization alignments $R_{\text{NL}}^{\uparrow/\downarrow}$ . Both curves exhibit antisymmetric behavior with respect to $H_\mathrm{x}$, suggesting that the \( y \) component of CSI, specifically unconventional CSI,\cite{32Yang2024-sa,33Chi2024-je,44Camosi2022-bu,45Ingla-Aynes2022-lw,46Ontoso2023-hj} does not exist in the Gr/Ta heterostructure system.

As $H_\mathrm{x}$ increases, $R_\mathrm{NL}$  converges between  $R_{\text{NL}}^\uparrow$ and $R_{\text{NL}}^\downarrow$ due to the magnetization pulling from the external magnetic field. However, a difference in $R_\mathrm{NL}$  values between the positive and negative sides is observed, resulting in an \( S \)-shaped feature in both curves. This behavior may originate from an \( x \)-direction component of CSI, such as the Rashba-Edelstein effect (REE),\cite{47Edelstein1990-rn, 08Manchon2015-uv} or it could be a spurious effect caused by the ordinary Hall effect in Gr due to the stray field of the FM electrode.~\cite{48Safeer2021-yr}

To investigate the origin of this \( S \)-shaped feature, additional non-local CSI measurements were conducted with a magnetic field applied along the \( z \)-axis (Note S10 ~\cite{supinfo}), inducing spin precession in the \( x \)-\( y \) plane. If the \( S \)-shaped curve resulted from the REE, antisymmetric precession curves would be expected since the Hanle precession curve is characteristic of CSI effects~\cite{48Safeer2021-yr}. However, the results do not show any precession and its absence is strong proof that the \( S \)-shaped background observed in Fig.~\ref{fig:CSI}b is an artifact rather than a genuine CSI effect.

Figure~4c displays the CSI resistance $R_\mathrm{CSI}$, the difference between the two curves in Fig.~4b ($R_\mathrm{CSI}=R_{\text{NL}}^\uparrow-R_{\text{NL}}^\downarrow)$, revealing a clear antisymmetric precession feature. This indicates the presence of a $z$-spin component of CSI in the Gr/Ta heterostructure, namely the SHE. To quantitatively analyze the SHE, we performed a numerical simulation to fit the antisymmetric precession curve (black curve in Fig.~4c) and extracted the $\theta_\mathrm{SH}$ using the spin transport parameters obtained from the Hanle precession measurements (see Note S7 for details \cite{supinfo}).

The gate voltage and temperature dependences of $\theta_\mathrm{SH}$ are presented in Figs.~4d and 4e, respectively. For the gate voltage dependence, the temperature was maintained at 50\,K, while for the temperature dependence, no gate voltage was applied. $\theta_\mathrm{SH}$ reaches its maximum value near $V_\mathrm{G}=0$ and diminishes as the magnitude of the gate voltage increases. This trend follows the variation of $R_\mathrm{sheet}$, suggesting that $\theta_\mathrm{SH}$ is largest when the Fermi level is near the CNP. Such behaviour is consistent with previously reported results for other Gr/metal~\cite{41Yang2023-ji} and Gr/TMD heterostructures.~\cite{34Safeer2019-bl, 35Benitez2020-mn}

\begin{table}[h!]
\centering
\caption{Comparison of CSI parameters in different Gr/metal oxide heterostructures}
\label{tab:csi_comparison}
\begin{ruledtabular}
\begin{tabular}{l r d d r l}
\textrm{System} & \textrm{$\lambda_s$ (nm)} & \multicolumn{1}{c}{\textrm{$\theta_\mathrm{SH}$ (\%)}} & \multicolumn{1}{c}{\textrm{$\lambda_\mathrm{CSI}$ (nm)}} & \textrm{T (K)} & \textrm{Ref.}\\
\colrule
Gr/TaO$_x$ & 530 & 1.2 & 6.3 & 50 & This Work \\
Gr/CuO$_x$ & 360 & 0.5 & 1.8 & 100 &~\cite{41Yang2023-ji} \\
Gr/Bi$_2$O$_3$ & 560 & 0.6 & 3.5 & 10 &~\cite{40Safeer2020-io} \\
\end{tabular}
\end{ruledtabular}
\end{table} 

As the temperature increases, $\theta_\mathrm{SH}$ decreases considerably. At room temperature, the signal becomes indistinguishable from the background noise (see Note S11~\cite{supinfo}). This reduction of CSI as temperature increases has been observed in a range of systems~\cite{35Benitez2020-mn, 36Herling2020-xi, 40Safeer2020-io, Yang2024-kv}, and can be attributed to a weakening of the proximity effect due to thermal smearing. To characterize the overall efficiency of CSI, we define a CSI efficiency, $\lambda_\mathrm{CSI}$, as the product of $\theta_\mathrm{SH}$ and  $\lambda_\mathrm{s}$ of the Gr/Ta heterostructure. Owing to the relatively large  $\lambda_\mathrm{s}$ over a range of temperatures, the Gr/Ta heterostructure achieves a notable $\lambda_\mathrm{CSI}$, reaching 6.3\,nm at 50\,K and 4.2\,nm at 150\,K (with no gate voltage). These values substantially exceed those reported for conventional SHE materials such as Pt (0.2\,nm@100\,K),\cite{11Rojas-Sanchez2014-wz} W (0.34\,nm@100\,K),\cite{49Kim2016-jm} and Cu/Au (0.17\,nm@100\,K).\cite{50Pham2021-bn} Moreover, they are comparable to the values observed in Gr/TMD heterostructures, such as Gr/WS$_2$ (3.75\,nm@300\,K)\cite{35Benitez2020-mn} and Gr/WSe$_2$ (4.9\,nm@300\,K).\cite{36Herling2020-xi} Given the more straightforward fabrication process of Gr/HM heterostructures compared to Gr/TMD systems, we propose that Gr/HM heterostructures can be considered promising candidates as efficient spin current sources for practical applications. A comparison with other Gr/metal oxide heterostructures is shown in Table I.

\begin{figure}[h!]
    \centering
    \includegraphics[width=0.48\textwidth]{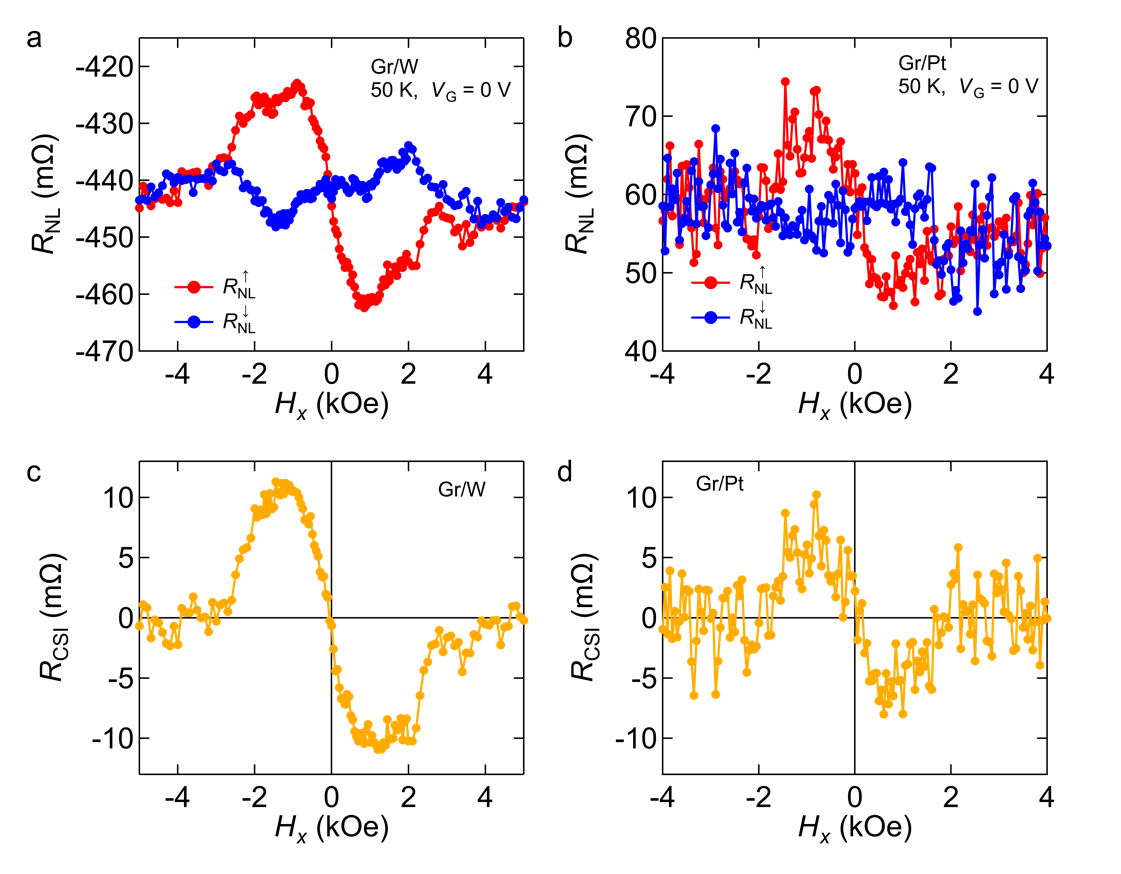}% Adjust figure path as needed
    \caption{\label{fig:Conclusion}
(a, b) Non-local resistance $R_{\text{NL}}$ for CSI measured as a function of $H_\mathrm{x}$ in Gr/W (a), and Gr/Pt (b) heterostructures. (c, d) The CSI resistance $R_{\mathrm{CSI}}$ extracted from the difference of the red and blue curves in (a) and (b) for Gr/W (c) and Gr/Pt (d) heterostructures. All the data were measured at 50 K without external gate voltages.
}
\end{figure}

To investigate the universality of the SHE in Gr/HM heterostructures, we replaced Ta with W and Pt and carried out the same non-local CSI measurements. The results are shown in Fig.~5. Figures~5a and~5b present the $H_x$-dependent $R_\mathrm{NL}^\mathrm{\uparrow/\downarrow}$ of CSI in the Gr/W and Gr/Pt heterostructures, respectively, measured at 50\,K with no applied gate voltage. In both cases, only an antisymmetric precession feature is observed, confirming the presence of the SHE in these heterostructures. To further support this conclusion, Figs.~5c and~5d display the CSI resistance $R_\mathrm{CSI}$ for Gr/W and Gr/Pt, respectively. For Gr/Ta we furthermore find similar results in a second device, as shown in Note S9~\cite{supinfo}. These results clearly demonstrate that the SHE can be consistently imprinted into Gr by sputtering a 5$d$ HM, underscoring the generality of this phenomenon in Gr/HM heterostructures.

Similarly to what has been discussed for graphene/CuO\(_x\)~\cite{41Yang2023-ji} and graphene/Bi$_2$O$_3$~\cite{40Safeer2020-io}, SOC can be induced by proximity because the Ta and W atoms within the oxide retain their strong intrinsic SOC. Orbital hybridization at the graphene/oxide interface is sufficient to imprint this strong SOC onto the graphene, leading to an intrinsic SHE. This mechanism is analogous to the well-documented proximity effect in Gr/TMD heterostructures. Alternatively, an extrinsic SHE can also be induced by adatom decoration, leading to skew scattering, a process that is also predicted theoretically~\cite{Ferreira2014-ct, Van_Tuan2016-vn}. However, experimentally distinguishing between the two mechanisms is not obvious, as previously discussed \cite{35Benitez2020-mn, 36Herling2020-xi}.

In the case of Pt, the CSI induced does not occur over the entire region, as in the case of Ta and W, but is instead localized to the regions in contact with the Pt, due to the discontinuous nature of the film. The mechanism of CSI for Pt could be due to either imprinted SOC or adatom-induced skew scattering, as discussed above. For Pt, there is the added complication that the metal remains conductive and, given its very short $\lambda_\mathrm{s}$, could contribute to spin absorption.

\section{Conclusion}

In summary, we have demonstrated a robust approach for realizing a strong SHE while preserving a long spin transport in Gr/HM heterostructures. By adopting an industry-friendly magnetron sputtering technique with grazing-incidence deposition, we effectively minimized structural damage to Gr while imprinting the HM's strong SOC. Our results show that the presence of the adjacent HM layer enhances $\theta_\mathrm{SH}$ in Gr with only a minor reduction in its spin diffusion length. Moreover, $\theta_\mathrm{SH}$ can be precisely tuned by applying a gate voltage, enabling a highly controllable CSI efficiency. Notably, the observed effect is consistent across different HMs, confirming the broad applicability of this method. These findings position Gr/HM heterostructures as a promising platform for large-scale spin current generation, offering both industrial feasibility and tunable functionality.

\begin{acknowledgments}
The authors acknowledge funding from MICIU/AEI/10.13039/501100011033 (Grant CEX2020-001038-M), from MICIU/AEI and ERDF/EU (Projects PID2021-122511OB-I00 and PID2021-128004NB-C21), from MICIU/AEI and the European Union NextGenerationEU/PRTR (Grant No. PCI2021-122038-2A), from the European Union's Horizon 2020 research and innovation programme under the Marie Skłodowska-Curie grant agreement No. 955671, and from the Valleytronics Intel Science and Technology Center. Z.C,  B.M.-G. and M.G. acknowledge funding from MICIU/AEI and the European Union NextGenerationEU/PRTR (Grant Nos. FJC2021-047257-I, RYC2021-034836-I and RYC2021-031705-I, respectively). H.Y. acknowledges support from the National Natural Science Foundation of China (Grant No. 62404014).
\end{acknowledgments}

\section*{Data Availability}

The data that support the findings of this study are openly available~\cite{zenodo_data_2024}.

\setlength{\parskip}{1.3em}
\setlength{\parindent}{0pt}
\raggedbottom

\renewcommand{\thesection}{S\arabic{section}}
\renewcommand{\thefigure}{S\arabic{figure}}
\renewcommand{\figurename}{Fig.}
\setcounter{section}{0}
\setcounter{figure}{0}

\clearpage
\mbox{}
\thispagestyle{empty}
\onecolumngrid

\begin{center}
    \textbf{\LARGE Supplementary Information}
\end{center}

\section{Formulation of the 3D FEM model}
\label{Formulation of the 3D FEM model}

As part of this work, we have developed a 3D finite element method (FEM) model to allow for the simulation of spin currents in our system, accounting for precession due to an external magnetic field. Analysis of proximitised graphene systems like those discussed here is typically carried out using a 1D analytical model for spin diffusion and spin-charge interconversion (SCI) \cite{safeer_room-temperature_2019xxx, herling_gate_2020xxx}. Previous works in modelling spin current using 3D finite element methods (FEM) have relied on a two‐channel model based on Valet-Fert theory\cite{Valet1993-hrxxx} in which one solves separately for the chemical potentials of spin‐up (\(\mu_{\uparrow}\)) and spin‐down (\(\mu_{\downarrow}\)) electrons\cite{Groen2021-isxxx, Pham2020-ruxxx, Yang2024-juxxx}. This approach will be discussed briefly, since it forms the basis of our approach to including spin precession in the graphene. Having solved for minority and majority spin potentials separately, we define the \emph{spin} and \emph{charge} chemical potentials as
\begin{equation} \label{eq:spin_charge_chemical_potential}
\mu_s = \frac{\mu_{\uparrow} - \mu_{\downarrow}}{2}, \quad \mu_c = \frac{\mu_{\uparrow} + \mu_{\downarrow}}{2}.
\end{equation}

The spin diffusion behaviour in a 3D system at equilibrium is given by \cite{Valet1993-hrxxx}:
\begin{equation} \label{eq:spin_diffusion_equation}
D_\mathrm{s} \nabla^2 \mu - \frac{\mu}{\tau_\mathrm{s}} = 0,
\end{equation}
where \(\mu\) can be the chemical potential of spin-up or spin-down electrons respectively, $D_\mathrm{s}$ is the spin diffusion coefficient, and $\tau_\mathrm{s}$ is the spin relaxation time.

Beginning from Eq. \ref{eq:spin_diffusion_equation}, and first to reduce the number of free variables, we divide through by $D_\mathrm{s}$ and introduce the term \(\sigma(\mathbf{r})\), the conductivity as a function of position, which accounts for the differing conductivity in each material, obtaining:
\begin{equation} \label{eq:diffusion_with_conductivity}
\nabla \cdot \left( \sigma(\mathbf{r}) \nabla \mu \right) - \frac{\sigma(\mathbf{r}) \mu}{D_\mathrm{s} \tau_\mathrm{s}} = 0.
\end{equation}

Using the relation \(\lambda_\mathrm{s}^2 = D_\mathrm{s} \tau_\mathrm{s}\), this becomes:
\begin{equation} \label{eq:diffusion_with_lambda}
\nabla \cdot \left( \sigma(\mathbf{r}) \nabla \mu \right) - \frac{\sigma(\mathbf{r}) \mu}{\lambda_\mathrm{s}^2} = 0.
\end{equation}
$\lambda_\mathrm{s}$ is typically the quantity of interest in spintronics, since the SCI efficiency is directly proportional to it \cite{Tao2018-zrxxx}, as visible in Eq.~\eqref{eq:DeltaRSCI}. Note that this is equivalent to Ohm's law with a relaxation term. Ohm's law can be written as:
\begin{equation} 
\label{eq:ohms_law}
\nabla \cdot \mathbf{J} = \nabla \cdot \left( \sigma(\mathbf{r}) \nabla \phi \right) = 0,
\end{equation}
in the more usual case of conservative current (\(\nabla \cdot \mathbf{J} = 0\)), where \(\phi\) is the electrical potential related to the charge chemical potential by \(\mu_\mathrm{c} = e\phi\). However, in this case, we are dealing with spin chemical potential, $\mu_\mathrm{s}$, rather than electrical potential, $\phi$, and therefore the equation includes a relaxation term proportional to \(1/\lambda_\mathrm{s}^2\), which accounts for spin relaxation effects. This relaxation term reflects the decay of the spin potential over a characteristic length scale \(\lambda_\mathrm{s}\).

\(\sigma(\mathbf{r})\) is considered to be isotropic in all cases, which is accurate for the FM. In the case of graphene, since the material is quasi-2D, the difference in in-plane and out-of-plane conductivity is assumed to have a negligible impact. The conductivity matrix \(\sigma(\mathbf{r})\) is therefore equal to the scalar conductivity times the identity matrix:
\begin{equation} \label{eq:isotropic_conductivity}
\sigma(\mathbf{r}) = \begin{pmatrix} \sigma & 0 & 0 \\ 0 & \sigma & 0 \\ 0 & 0 & \sigma \end{pmatrix},
\end{equation}
where \(\sigma\) is the scalar conductivity.

This isotropic conductivity tensor does not hold in regions where we are considering spin-charge interconversion (SCI). In this system, we have the spin Hall effect (SHE), which we implement via off-diagonal terms in the conductivity matrix. For spins polarized along \(y\), the spin Hall conductivity \(\sigma_\mathrm{SH}(\mathbf{r})\) becomes spin-dependent. For spin \(y \uparrow\), with spin current along \(x\), the conductivity tensor is:
\begin{equation} \label{eq:she_y_up}
\sigma_\mathrm{SH}(\mathbf{r})_{y\uparrow} = \sigma \cdot 
\begin{pmatrix} 
1 & -\theta_\mathrm{SH} & 0 \\ 
\theta_\mathrm{SH} & 1 & 0 \\ 
0 & 0 & 1 
\end{pmatrix},
\end{equation}
where \(\theta_\mathrm{SH}\) is the spin Hall angle, and the off-diagonal terms represent the coupling between charge and spin currents due to the spin Hall effect. For spin \(y \downarrow\), the spin Hall contributions have opposite sign, and the conductivity tensor is:
\begin{equation} \label{eq:she_y_down}
\sigma_\mathrm{SH}(\mathbf{r})_{y\downarrow} = \sigma \cdot 
\begin{pmatrix} 
1 & \theta_\mathrm{SH} & 0 \\ 
-\theta_\mathrm{SH} & 1 & 0 \\ 
0 & 0 & 1 
\end{pmatrix}.
\end{equation}

Different effects can be implemented in the model via appropriate modification of the conductivity tensor, such as the anomalous Hall effect. To solve Eq.~\eqref{eq:diffusion_with_lambda} via FEM, we need to convert this analytical equation to its corresponding weak form. 
Starting with Eq. \ref{eq:diffusion_with_lambda}
we multiply through by the test function \(v(\mathbf{r})\) and integrate over the volume \(\Omega\), yielding:
\begin{equation} \label{eq:weak_form}
\int_{\Omega} v(\mathbf{r}) \left( \nabla \cdot \left( \sigma(\mathbf{r}) \nabla \mu \right) - \frac{\sigma(\mathbf{r}) \mu}{\lambda_\mathrm{s}^2} \right) \, dV = 0.
\end{equation}

Using the divergence theorem, the first term becomes:
\begin{equation} \label{eq:divergence_theorem}
\int_{\Omega} v(\mathbf{r}) \nabla \cdot \left( \sigma(\mathbf{r}) \nabla \mu \right) \, dV = - \int_{\Omega} \sigma(\mathbf{r}) \nabla \mu \cdot \nabla v(\mathbf{r}) \, dV + \int_{S} v(\mathbf{r}) \sigma(\mathbf{r}) \nabla \mu \cdot \mathbf{n} \, dS.
\end{equation}
Substituting this back, the weak form of the equation is:
\begin{equation} \label{eq:weak_form_substituted}
- \int_{\Omega} \sigma(\mathbf{r}) \nabla \mu \cdot \nabla v(\mathbf{r}) \, dV + \int_{S} v(\mathbf{r}) \sigma(\mathbf{r}) \nabla \mu \cdot \mathbf{n} \, dS + \int_{\Omega} \frac{\sigma(\mathbf{r}) \mu}{\lambda_\mathrm{s}^2} v(\mathbf{r}) \, dV = 0.
\end{equation}

Here, \(v(\mathbf{r})\) is the test function, \(\Omega\) is the volume over which we are solving, $S$ is the surface, and \(\mathbf{n}\) is the outward unit normal vector on $S$. This is the equation we solve using FEM and the Galerkin method, solving it separately for each spin channel. The second term here corresponds to the surfaces of the 3D system and contains the boundary conditions applied to the model. We apply Neumann boundary conditions corresponding to the applied current according to the second term in Eq. \ref{eq:weak_form_substituted}, expressed as:

\begin{equation} \label{eq:neumann_boundary_condition}
\nabla \mu_\mathrm{c} = \rho \cdot \frac{I_c}{A},
\end{equation}

where \(\mu_\mathrm{c}\) is the charge chemical potential, \(\rho\) is the resistivity, \(I_c\) is the charge current, and \(A\) is the area of the surface.

The process can be simplified since all the values involve a non-local resistance, and thus the results are normalized by the input current. Without loss of generality, we set \(I_c = 1 \mathrm{A}\) throughout, so that aside from units the measured voltage is equal to the resistance. To determine the voltage on a surface, which corresponds to the experimentally measured value, we use:

\begin{equation} \label{eq:voltage_surface}
V = \frac{\int \mu_c \, dS}{A},
\end{equation}

where \(V\) is the voltage, \(\mu_c\) is the charge chemical potential, and \(dS\) is a surface element. To simulate the experimental measurement, we find the voltage on the two faces corresponding to $V$+ and $V$- in the experiment, and these are printed to the simulation output.

The model consists of three regions: the pristine graphene channel (Gr), the proximitized graphene (Gr/Ta), and the FM electrodes. In the case of Gr and FM, the regions are connected by an interface, which couples the chemical potentials on each side. While the dimensions of the FM and graphene are precisely those derived from SEM, the interface is assumed to be truly 2D. Experimentally, we deposit 0.3 nm of Ti, which is allowed to oxidize, meaning the interface has some finite thickness, although it is difficult to precisely determine. However, in practice, it is the resistance of the interface that is important, so we implement it in the model as a 2D surface between the Gr and FM, with an effective resistance corresponding to the experimental value. This is equivalent to using a small volume with a corresponding resistivity but is computationally much easier.

The spin-dependent conductivity at the interface follows the Valet-Fert model for each channel, where the conductivity of the up and down spins is described by:
\begin{equation} \label{eq:valet_fert_interface}
\sigma_{\uparrow} = \frac{\sigma_i}{2} (1 + \gamma m),
\end{equation}
\begin{equation} \label{eq:valet_fert_interface_2}
\sigma_{\downarrow} = \frac{\sigma_i}{2} (1 - \gamma m),
\end{equation}
where \(\gamma\) is the interface polarization term, \(m\) is the FM magnetization (equal to \(\pm 1\)) in this two-channel model, and \(\sigma_i\) is the interface conductivity.

The interface conductivity is defined as:
\begin{equation} \label{eq:interface_conductivity}
\sigma_i = \frac{1}{R_\mathrm{i} A_\mathrm{i}},
\end{equation}
where \(R_i\) is the experimentally measured interface resistance and \(A_\mathrm{i}\) is the interface area. Given the highly resistive interfaces involved, this polarization dominates and corresponds to \(P_\mathrm{i}\) in Eq.~\eqref{eq:DeltaRNL}. Although the model supports a bulk polarization for the FM, this value does not matter in practice due to the overwhelming influence of the interface polarization, given its high resistance. In systems with more transparent interfaces, the bulk polarization becomes relevant. In such cases, the bulk spin-dependent conductivity is defined as:
\begin{equation} \label{eq:bulk_spin_conductivity}
\sigma_{\uparrow} = \frac{\sigma}{2} (1 + \beta m),
\end{equation}
\begin{equation} \label{eq:bulk_spin_conductivity_2}
\sigma_{\downarrow} = \frac{\sigma}{2} (1 - \beta m),
\end{equation}
where \(\beta\) is the bulk polarization term and \(\sigma\) is the bulk conductivity.

Each physical region (Gr, FM, interfaces, Gr/Ta) has different associated physical quantities and equations. For example, all regions have an associated conductivity for each spin channel, but this is different in each region. In the FM, there is a connection between the conductivity and the magnetisation, but in Gr there is no such relationship. Furthermore, the spin diffusion parameters are defined separately in each region, with $\lambda_\mathrm{s}$ being significantly larger in the Gr than in the FM.

A similar, albeit more complex, model can be applied to systems with the introduction of spin precession in the graphene. The spin diffusion equation now contains a precession term and is given by\cite{Fabian2007-isxxx}:
\begin{equation} \label{eq:spin_diffusion_precession}
D_\mathrm{s} \frac{\partial^2 \mu}{\partial r^2} - \frac{\mu}{\tau_\mathrm{s}} + \omega \times \mu = 0,
\end{equation}
This additional term, \(\omega \times \mu\), where $\omega = \frac{g \mu_B H_\mathrm{z}}{\hbar}$ is the Larmor frequency, with $g$ being the Landé $g$-factor, $\mu_B$ the Bohr magneton, and $\hbar$ the reduced Planck's constant, represents the spin precession due to an external or effective magnetic field, making the behaviour of the system more complex compared to the purely diffusive case.

This introduces several complications to the model. Firstly, there is now a coupling between the \(y\) and \(z\) components of the spin chemical potential, given by the \(\omega \times \mu\) term in the equation (and since we apply $H_\mathrm{x}$). As a result, the two-channel model is no longer sufficient. Secondly, the description of the FM magnetization as being \(\pm 1\) is no longer adequate, and we need to consider the full range of magnetization in the \(xy\)-plane. This is further complicated by the pulling of the magnetisation by the applied field, which means that not only does the field affect the spin behaviour, it also affects the magnetization of the FM elements. In this work, we only consider an applied field along \(x\), which simplifies computation. However, for generality, we construct the simulation to have six channels, corresponding to \(\mu^\uparrow_x\), \(\mu^\uparrow_y\), \(\mu^\uparrow_z\), \(\mu^\downarrow_x\), \(\mu^\downarrow_y\), and \(\mu^\downarrow_z\).

The coupling between \(\mu^y\) and \(\mu^z\) is included by expanding the cross-product term \(\omega \times \mu\). The contribution to \(\mu^y_\uparrow\) from \(\mu^z_\uparrow\) due to \(H_\mathrm{x}\) is given, for example, by:
\begin{equation} \label{eq:coupling_precession}
D_\mathrm{s} \frac{\partial^2 \mu^y_\uparrow}{\partial r^2} - \frac{\mu^y_\uparrow}{\tau_\mathrm{s}} + \omega \mu^z_\uparrow = 0,
\end{equation}This same process is repeated for all couplings induced by the magnetic field. For inclusion of these terms in the weak form, we note that they are mathematically equivalent to the spin relaxation term, with the sign depending on the direction of precession. Therefore a similar derivation is applied, with the resultant equations being different for each spin channel.

Similarly to Eq. \eqref{eq:spin_charge_chemical_potential}, we can define the spin and charge chemical chemical potentials in terms of the six spin channels. The spin potential now contains three components, defined as:

\begin{equation} \label{eq:spin_potentials}
\mu_s^x = \frac{\mu_\uparrow^x - \mu_\downarrow^x}{2}, \quad 
\mu_s^y = \frac{\mu_\uparrow^y - \mu_\downarrow^y}{2}, \quad 
\mu_s^z = \frac{\mu_\uparrow^z - \mu_\downarrow^z}{2}.
\end{equation}

The charge potential is then defined as:

\begin{equation} \label{eq:charge_potential}
\mu_c = \frac{\mu_\uparrow^x + \mu_\uparrow^y + \mu_\uparrow^z + \mu_\downarrow^x + \mu_\downarrow^y + \mu_\downarrow^z}{6}.
\end{equation}

Without considering precession, the quantities $D_\mathrm{s}$ and $\tau_\mathrm{s}$ only ever appear as a product, so it is sufficient to consider only \(\lambda_\mathrm{s}\). However, the precession term means that $D_\mathrm{s}$ now appears separately from $\tau_\mathrm{s}$, so we need to consider both separately and calculate the value of \(\lambda_\mathrm{s}\) afterward. Like \(\sigma\), both $D_\mathrm{s}$ and $\tau_\mathrm{s}$ vary across different regions of the system. In the FM, there is no precession due to the external field, so it is sufficient to define the relaxation terms from Eq.~\eqref{eq:diffusion_with_lambda} only in terms of \(\lambda_\mathrm{s}\). In the FM, \(\lambda_\mathrm{s}\) is approximately 5 nm, the precise value does not affect the outcome due to the very large interface resistance which dominates the behaviour. In the graphene region, we need to separately define the values for $D_\mathrm{s}$, $\tau_\mathrm{s}$, and \(\rho\). Likewise, in the Gr/Ta region, different values for these three quantities must be defined to account for the material-specific properties of spin diffusion and relaxation.

The weak forms of the equations, along with the boundary conditions, are input in the open source finite element solver GetDP\cite{getdpxxx}, along with the associated 3D volume discussed in the next section, which allows for an accurate simulation of the spin and charge chemical potentials throughout the system.

\subsection{Comparison of 1D and 3D approaches in ideal circumstances}

To ensure the accuracy of the 3D model, we can apply it to a very simple system, consisting of a perfectly rectangular piece of graphene with homogeneous properties, where we apply a current of 1 A from the left boundary to the right boundary. The current has a spin polarization of 10\%, and we set the graphene volume to have a square resistance of 1000 $\Omega/sq$, a $\tau_\mathrm{s}$ of 50 ps, and a $D_\mathrm{s}$ of $0.2 \times 10^{16}$ nm\(^2\)/s. This system has an analytical solution, given by the equation\cite{safeer_room-temperature_2019xxx}:

\begin{equation}
\Delta \mathit{R}_{\mathrm{NL}}
= \frac{P^2 \cos^2(\beta)\, \mathit{R}_{\mathrm{sq}} \,\lambda_{\mathrm{s}}}{w}
  \,\mathcal{R} \Biggl\{
    \frac{e^{-\frac{L}{\lambda_{\mathrm{s}}}\,\Omega}}{\Omega}
  \Biggr\},
\label{eq:DeltaRNL}
\end{equation}

where $\Delta \mathit{R}_{\mathrm{NL}} = (\mathit{R}_{\mathrm{NL}}^{\mathrm{P}} - \mathit{R}_{\mathrm{NL}}^{\mathrm{AP}})$ is the pure spin precession (Hanle) signal. $\tau_{\mathrm{s}}$ is the spin lifetime, $\mathit{D}_{\mathrm{s}}$ is the spin diffusion coefficient, and $\lambda_{\mathrm{s}} = \sqrt{\mathit{D}_{\mathrm{s}} \tau_{\mathrm{s}}}$ is the spin diffusion length. $\mathit{P}$, is the spin polarization of the current. $\mathit{R}_{\mathrm{sq}}$ is the square resistance of graphene, $w$ is the channel width, $L$ is the channel length, and $\Omega = \sqrt{1 - i \omega \tau_{\mathrm{s}}}$. $\beta$ is the angle of the Co magnetization with respect to its easy axis, which for this analysis is $\pi/2$, but in real devices varies with the applied field as discussed later (See Note \ref{Magnetisation pulling}).

\begin{figure}[htpb]
    \centering
    \includegraphics[width=1\linewidth]
{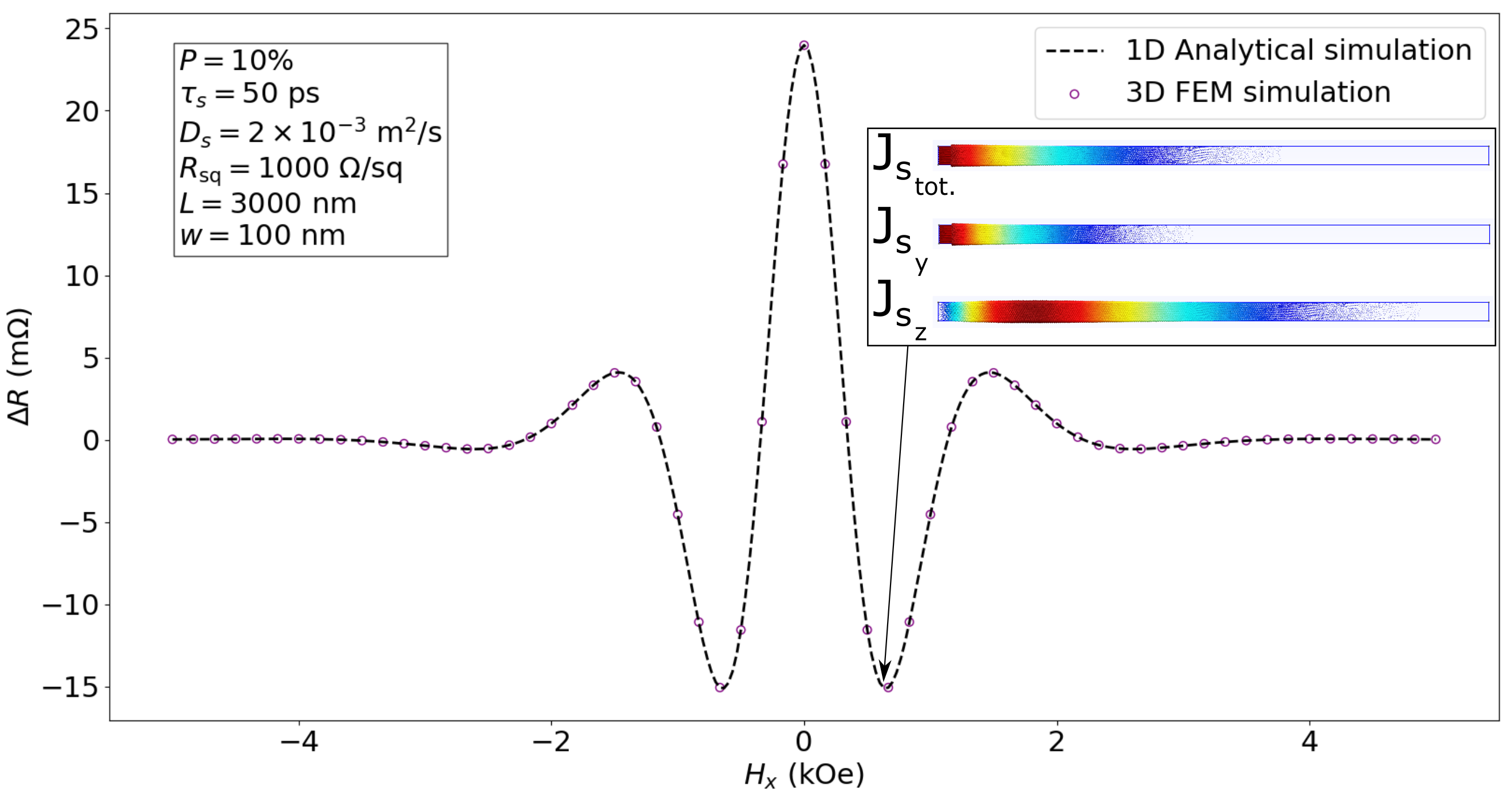}
\caption{Comparison of the 1D and 3D models for spin diffusion compared under ideal conditions and using the same parameters. Inset are the total spin current, distributed along the channel (x-direction), as well as the contributions thereto of the z and y polarised spins for a spin current initially polarised fully in y at the left side of the channel, at an applied field of 0.68 kOe}
    % \captionsetup{justification=justified}
    \label{fig:1D3DIdeal}
\end{figure}

The results are shown in Fig.~\ref{fig:1D3DIdeal}, indicating perfect agreement between the two models in this ideal scenario.

\FloatBarrier

Similarly, for SCI in a perfect Hall bar geometry we have an analytical solution given by:
\cite{safeer_room-temperature_2019xxx}
\begin{equation}
\mathit{R}_{\mathrm{SCI}} 
= \frac{P\,\cos(\beta)\,\theta_{\mathrm{SH}} \, \mathit{R}_{\mathrm{sq}} \,\lambda_{\mathrm{s}}}{w_{\mathrm{cr}}} 
\, \mathrm{Im} \Biggl\{
\frac{ 
   e^{-\frac{L}{\lambda_{\mathrm{s}}}\,\Omega}
   \;-\;
   e^{-\frac{(L + w_{\mathrm{cr}})}{\lambda_{\mathrm{s}}}\,\Omega}
}
{\Omega}
\Biggr\},
\label{eq:DeltaRSCI}
\end{equation}

where \(w_{\mathrm{cr}}\) is the proximitized channel width. The efficiency of spin-charge interconversion, \(\theta_{\mathrm{SH}}\), is defined as \( \mathit{J}_\mathrm{s} / \mathit{J}_{\mathrm{c}} \). The other symbols have the same meaning as in Eq.~\ref{eq:DeltaRNL}.

\begin{figure}[htpb]
    \centering
    \includegraphics[width=1\linewidth]{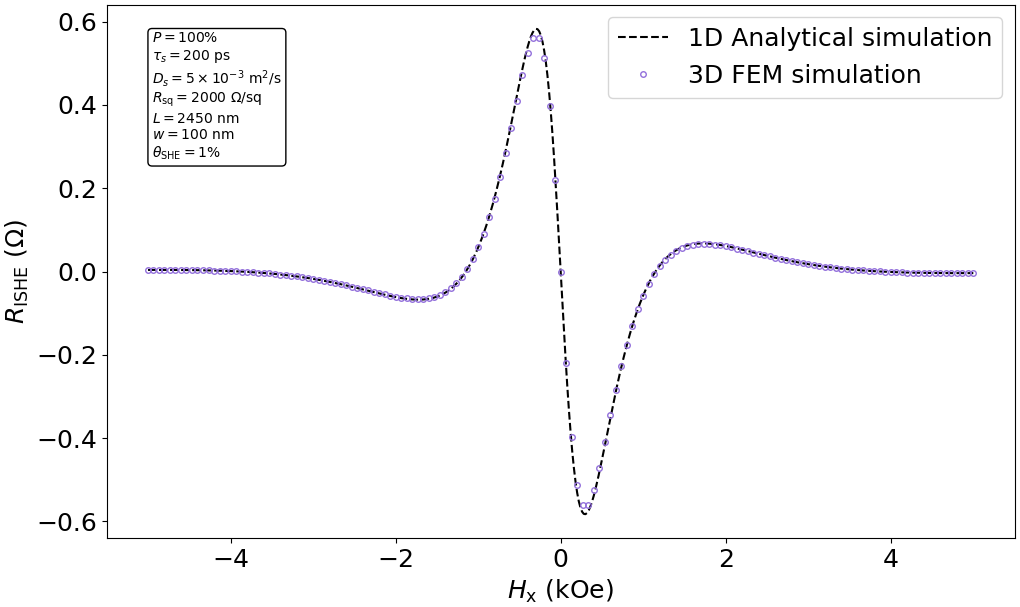}
\caption{Comparison of the 1D and 3D models for SCI compared under ideal conditions and using the same parameters in a Hall bar geometry.}
    \label{fig:SCC1D3DIdeal}
\end{figure}

We solve Eq.~\ref{eq:DeltaRSCI} for the arbitrarily chosen values 
\(\lambda_{\mathrm{s}} = 1000\,\mathrm{nm}\), 
\(P = 1\), 
\(\theta_{\mathrm{SH}} = 0.01\), 
\(R_{\mathrm{sq}} = 2000\,\Omega /\square\), 
and \(w_{\mathrm{cr}} = 100\,\mathrm{nm}\). 
We perform the same calculation using the 3D FEM simulation, with the outputs shown in Fig.~\ref{fig:SCC1D3DIdeal}.

\FloatBarrier
\clearpage
\section{3D Geometry}

To extract the dimensions of the device, we use SEM to produce a highly accurate map of the device, showing both the pristine and proximitized channels, as discussed in the main text. These dimensions are then used to construct an accurate model of the system. The key deviations from the 1D approach are the finite width of the FM contacts, the slight rotation of the channel relative to the FM (away from \(90^\circ\)), and the slightly off-centre deposition of the Ta. To create the 3D geometry, we use these dimensions along with the open-source software Gmsh \cite{Geuzaine2009-ufxxx} to generate the finite element mesh over which our solution functions are calculated.

\begin{figure}[htpb]
    \centering
    \includegraphics[width=0.8\linewidth]{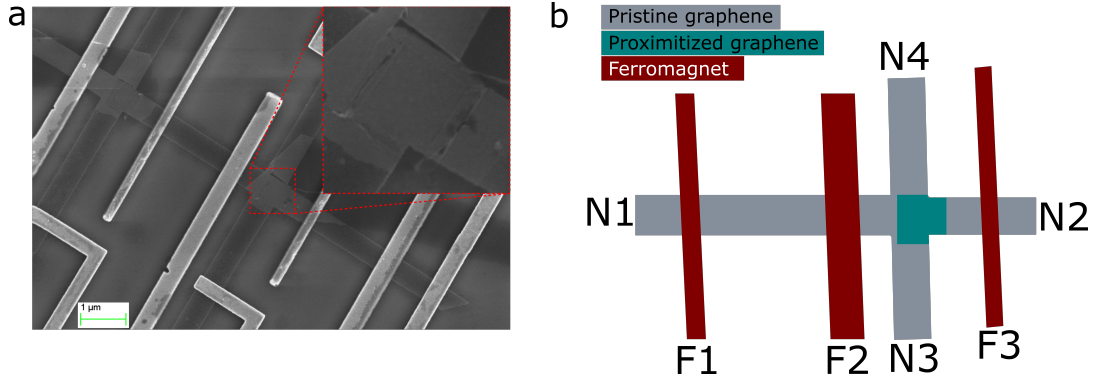}
    \caption{(a) SEM image of the device showing the graphene Hall bar, FM electrodes, and the region proximitized by TaO\textsubscript{x}. Inset: zoom of the proximitised area (b) Schematic showing the different regions considered in the model with dimensions derived from the SEM pictures. The labels correspond to the contacts in Fig. 2 of the main text.}
    \label{fig:SEM}
\end{figure}

\FloatBarrier
\clearpage
\section{Resistivity}
\label{Resistivity}
The first step is to use the 3D model to calculate the resistivity of each region. This can be done relatively easily since the resistivity is a property of electrical current, with the spin resistivity assumed to be the same as that for electrical current in the graphene. Most of the complexity discussed in Note \ref{Formulation of the 3D FEM model} can therefore be discarded, and it is sufficient to simulate just one conduction channel for a conservative electrical current. Experimentally, the pristine region is measured in a 4-point configuration, with current applied from \(N_2\) to \(N_1\) and voltage measured between \(F_2\) and \(F_1\). This resistance can be used to calculate the resistivity \(\rho\) and square resistance $R_\mathrm{sq}$ analytically, since the channel is a well-defined shape. As confirmation, the same measurement is simulated using the 3D approach and the SEM dimensions, and the values agree to within 5\%.

The resistivity \(\rho\) is calculated as:
\begin{equation}
    \rho = \frac{R \cdot w \cdot t}{L},
    \label{eq:resistivity}
\end{equation}
where \(R\) is the measured resistance, $W$ is the width of the channel, \(t\) is the thickness of the graphene, and $L$ is the distance between the voltage probes (measured from the centers of the ferromagnets).

The square resistance $R_\mathrm{sq}$ is given by:
\begin{equation}
    R_{sq} = \frac{\rho}{t} = \frac{R \cdot w}{L}.
    \label{eq:sheet_resistance}
\end{equation}

$R_\mathrm{sq}$ is used because it is independent of the thickness of the graphene and can be used even in cases where the number of layers is unknown. This is found as the resistance multiplied by the aspect ratio in the case of the Gr measurement, but is more complex for the Gr/Ta data, since the width varies due to the arms of the Hall bar. For the 3D model, we need to input resistivity values in each region, not $R_\mathrm{sq}$ values, since all elements, including the Gr, have some finite thickness. However, for unit (1 nm) thick graphene, the two are equivalent (Eq. \ref{eq:sheet_resistance}). In the 2D limit, changing the thickness of the graphene in the model has almost no effect, provided the corresponding resistivity \(\rho\) that gives the same $R_\mathrm{sq}$ is used. Although we do not have highly accurate thickness data for the Gr, we confirm the thickness independence of the results for thicknesses between 1 and 10 monolayers, and then for numerical convenience use a thickness of 1 nm (corresponding to around 3 monolayers), since the base unit of the model is nm this means \(R_{sq}=\rho\).

\begin{figure}[htpb]
    \centering
    \includegraphics[width=1\linewidth]{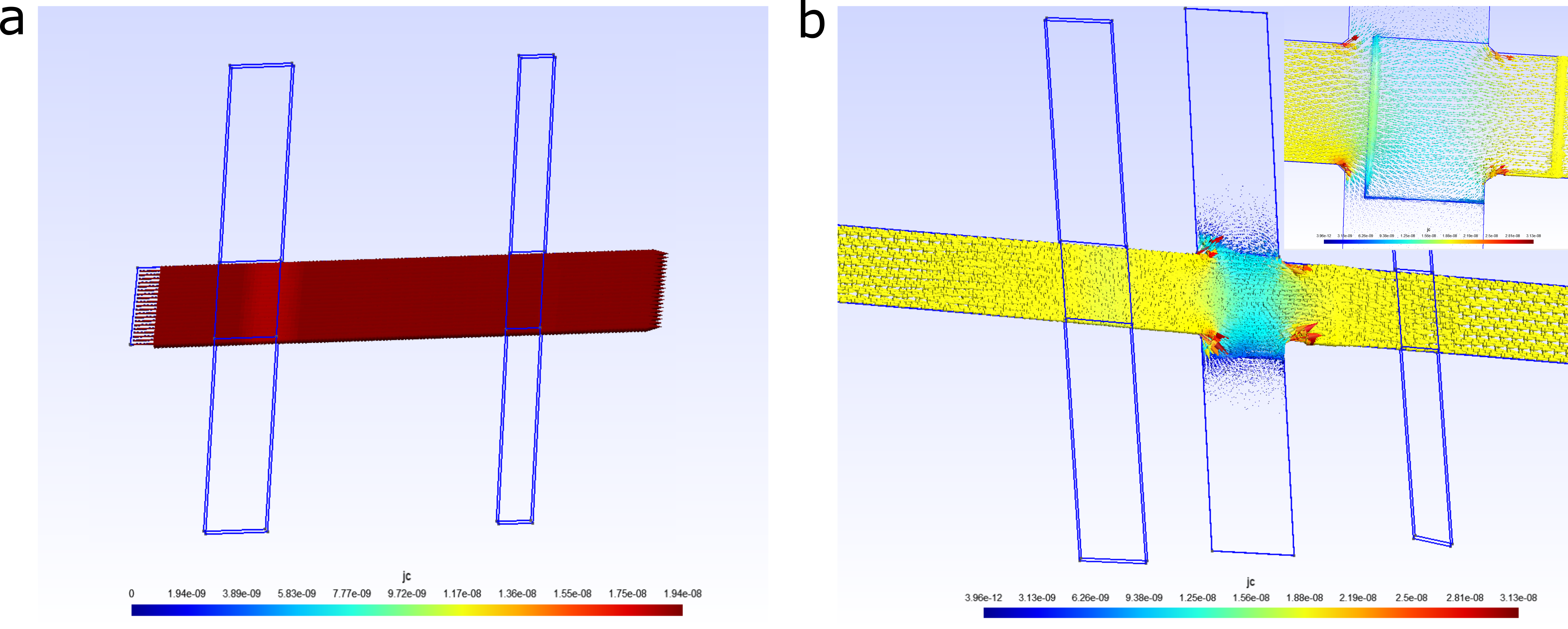}
    \caption{(a) Screenshot of the 3D simulation showing the pristine Gr channel, with the charge current density displayed. Also shown are the two FM contacts (b) Charge current density, now shown through the Gr/Ta part of the device, indicating the small deviation from linear current through the centre of the device. The inset shows the central region for increased clarity. The Gr and Gr/Ta regions have different values of $\rho$ but they are similar enough not to be visible visually in this simulation.}
    \label{fig:rhosimscreenshot}
\end{figure}

We can repeat this analysis using the 3D model (Fig. \ref{fig:rhosimscreenshot}a) and obtain almost exactly the same results. However, for the crossing measurement, the situation is more complicated. There are two main issues, firstly, the geometry is no longer 1D, making the application of Eq. \eqref{eq:sheet_resistance} more difficult, and secondly, the measurement contains contributions from both the Gr and Gr/Ta regions.

\begin{figure}[htpb]
    \centering
    \includegraphics[width=1\linewidth]{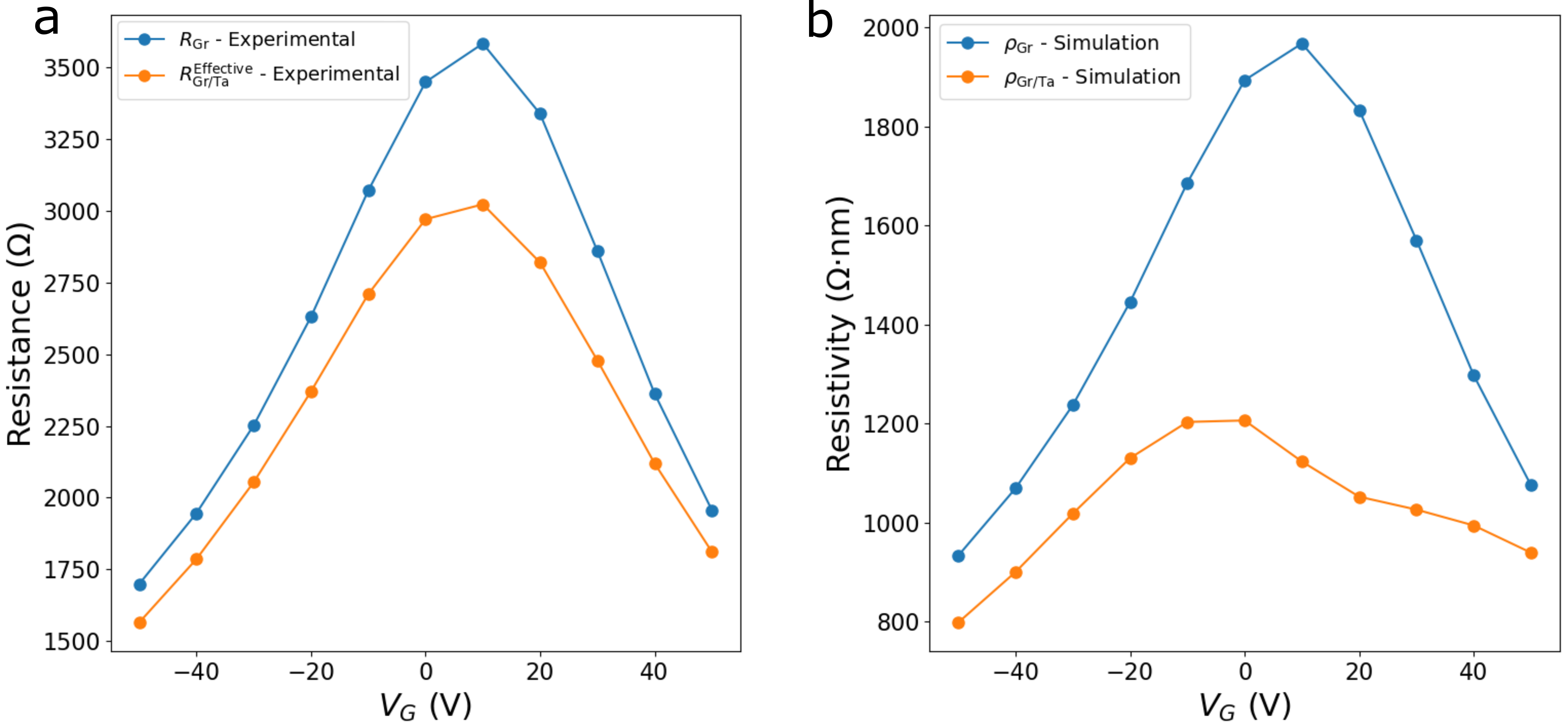}
    \caption{(a) The experimental resistance values for the Gr and effective values of the Gr/Ta region as a function of back-gate voltage ($V_\mathrm{G}$) measured at 50\,K. This corresponds to Fig. 1c in the main text. (b) Simulation results showing the resistivity for each region, where we have disentangled the contributions of the Gr and Gr/Ta regions to demonstrate that the deposited metal has a large impact on the Gr resistivity.}
    \label{fig:rhosimresults}
\end{figure}

The data shown in Fig. \ref{fig:rhosimresults} represents an effective square resistance for the entire mixed region. This situation could be approximated reasonably well as a series resistor mode. However, to accurately isolate the resistance of the Gr/Ta square, we simulate the system using our 3D model.

First, we determine \(\rho_{\text{Gr}}\) by simulating the linear pristine channel (Fig. \ref{fig:rhosimscreenshot}a). Then, we use this value to fix the resistivity, \(\rho\), in the Gr part of the Gr/Ta measurement. Finally, we run the simulation shown in Fig. \ref{fig:rhosimscreenshot}b, varying the value of \(\rho_{\text{Gr/Ta}}\) until the simulated output voltage matches the experimental value. This process is repeated for all values of back-gate voltage, \(V_G\), as summarized in Fig. \ref{fig:rhosimresults}.

\FloatBarrier

\clearpage
\section{Magnetization pulling}
\label{Magnetisation pulling}
In this work, we study the SHE, which depends on the \(z\)-polarized component of the spin, generated from the precession of the injected \(y\)-component. We apply the external magnetic field along the \(x\)-direction, causing spin precession. However, as a secondary effect, this field also pulls the magnetization of the FMs away from the \(y\)-axis and into the \(x\)-axis. This is a significant effect, as the magnetization of the injecting electrode saturates along the \(x\)-direction at around 0.3 T, which lies within the range of our spin transport measurements. To accurately study the system, we need to extract an expression for the magnetization as a function of the applied field so this can be input into the model along with the field value. 

\begin{figure}[htpb]
    \centering
    \includegraphics[width=1\linewidth]{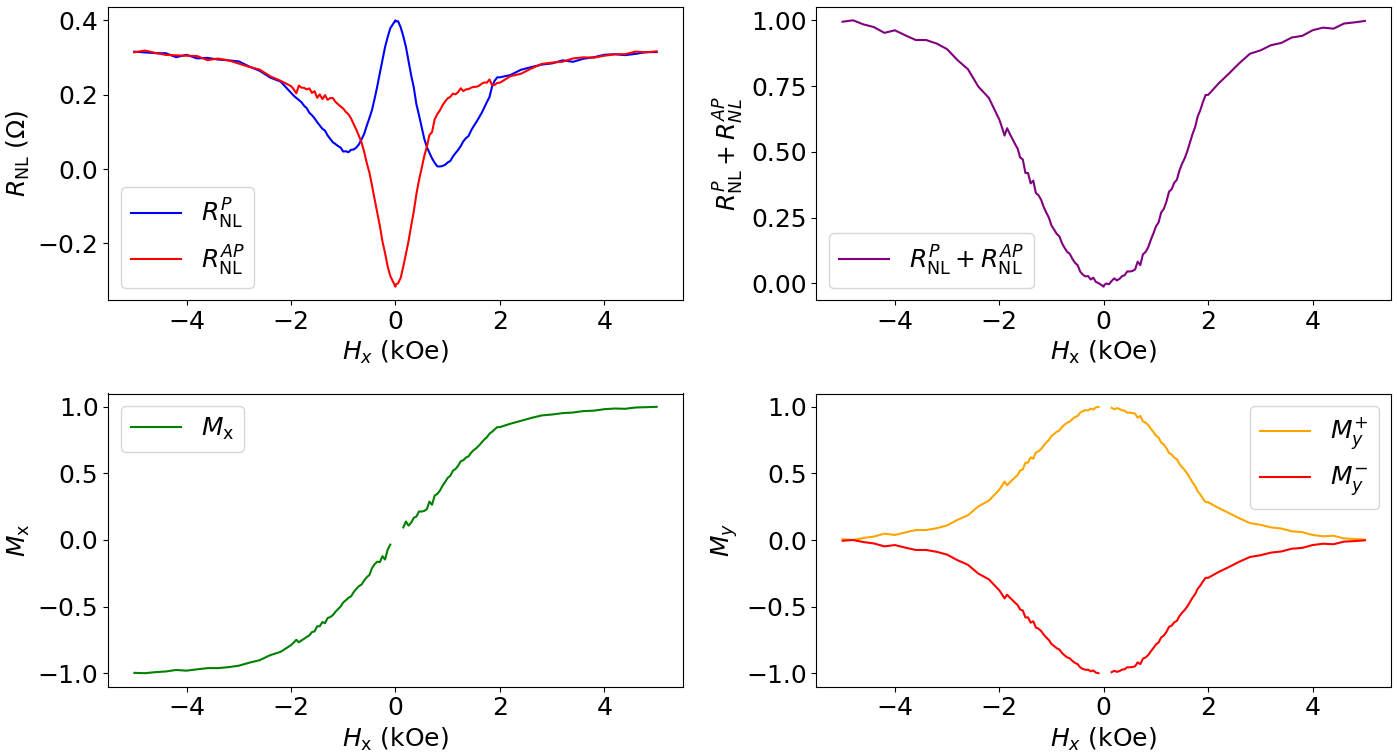}
    \caption{(a) The Hanle precession curves at 50 K and $V_G=0$ in pristine Gr showing the measured non-local resistance for both parallel ($R_\mathrm{NL}^\mathrm{P}$) and anti-parallel($R_\mathrm{NL}^\mathrm{AP}$) configurations of injecting and detecting FM contacts (b) The sum of the two curves in (a) normalized between 0 and 1. (c) The x component of the unit magnetisation vector as a function of $H_\mathrm{x}$. (d) The corresponding y-component, for initial magnetisation along $\pm$ y respectively}
    \label{fig:Magnetisation}
\end{figure}

To do this, we note that when we subtract \(R_{\mathrm{NL}}^{\mathrm{AP}}\) from \(R_{\mathrm{NL}}^{\mathrm{P}}\), we remove the contribution of this pulling effect, isolating only the precession, since the contribution of the contact pulling is the same in each measurement, but the spin precession component changes sign. Conversely, if we instead add the two measurements together, we remove the precession component, isolating the contribution from the magnetization pulling. This is shown in Fig. \ref{fig:Magnetisation}c, where we can see the initial rotation followed by saturation of the magnetization. 

The spin transport measurement depends on both the injector and detector magnetization and therefore contains a factor proportional to \(M_x^2\)—this corresponds to the \(\cos^2(\beta)\) term in Eq. \eqref{eq:DeltaRNL}. To extract the magnetization, we take the spin transport signals from Fig. \ref{fig:Magnetisation}a, add them together, normalize the result from 0 to 1 (Fig. \ref{fig:Magnetisation}b). We then take the square root and multiply this value by the sign of the field, since this affects the direction of pulling, which directly yields the \(x\)-component of the magnetization vector \(M_x\) (Fig. \ref{fig:Magnetisation}c). This is a normalized vector satisfying \(M_x^2 + M_y^2 = 1\), thus also giving us \(M_y\) (Fig. \ref{fig:Magnetisation}d). 

The magnetization values associated with each applied field are found and put into the 3D model in the FM regions. The curve is smoothly interpolated to remove the gap around \(H_\mathrm{x} = 0\). The value of \(M_y\) depends on the initial magnetization of the FM, but \(M_x\) depends only on the field.

\FloatBarrier
\clearpage
\section{Spin precession in the pristine graphene}

Having established the resistivity (Note \ref{Resistivity}) in the pristine graphene, and the magnetisation pulling (Note \ref{Magnetisation pulling}) we can now aim to extract the spin transport parameters \(\displaystyle D_{s}\), \(\displaystyle \tau_{s}\), and \(\displaystyle \lambda_{s}\). Since these values are connected, we can perform all the fitting with \(\displaystyle D_\mathrm{s}\) and \(\displaystyle \tau_\mathrm{s}\), with \(\displaystyle \lambda_{s}\) being a derived quantity, although the choice of which two parameters to use is arbitrary.

\begin{figure}[htpb]
    \centering
    \includegraphics[width=0.7\linewidth]{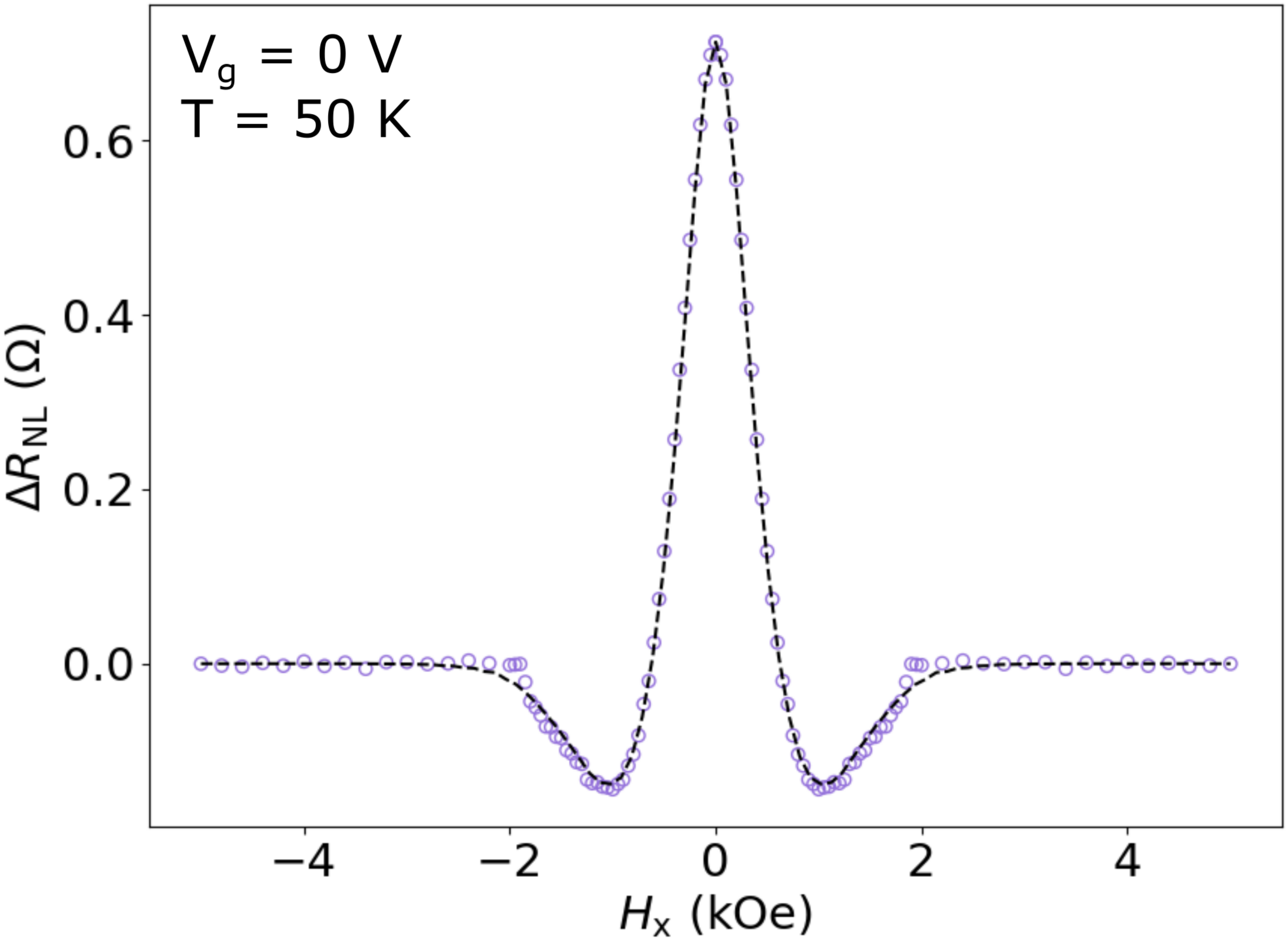}
    \caption{Symmetrised Hanle precession signal of pristine Gr at 50\,K (open circles) with the results of 3D fitting shown as a black dashed line.}
    \label{fig:HanlePristineSimFit}
\end{figure}

We begin with the parallel~(P) and anti-parallel~(AP) measurements, and subtract them to find $\Delta R_\mathrm{NL}$, as shown in Fig. \ref{fig:HanlePristineSimFit}. This data corresponds to current injection from \(N_1\) to \(F_1\), and voltage measurement from \(N_2\) to \(F_2\) (Fig. \ref{fig:SEM}b). In the simulation, we apply boundary conditions corresponding to this applied current (Eq. \ref{eq:neumann_boundary_condition}), and on completion of the simulation we measure the voltage on the two faces (Eq. \ref{eq:voltage_surface}), subtract them, and divide by the applied current to get a simulated value of \(R_\mathrm{NL}\). This is repeated for both the P and AP cases, and the values are subtracted to leave just the simulated spin precession signal.

This simulation needs to be repeated for each value of \(H_\mathrm{x}\), but in practice, this is computationally very expensive. Therefore, a discrete selection of field values are used to do the fitting. Having finished this simulation, we obtain a simulated spin diffusion curve (Fig. \ref{fig:HanlePristineSimFit}), corresponding to a value of \(\displaystyle D_\mathrm{s}\) and a value of \(\displaystyle \tau_\mathrm{s}\). We compare this to the experimental measurement, however to reduce computation time we symmetrise the experimental data first (Fig. \ref{fig:HanlePristineSimFit}), since there are minor differences in the positive and negative field $R_\mathrm{NL}$ values for experimental reasons (Fig. 2f, main text). These differences are obviously not symmetric so cannot be modelled by our symmetric model, and are in any case rather minor. Therefore, by using the symmetrised data we can just fit half the data, with the fit (Fig. \ref{fig:HanlePristineSimFit}) then corresponding to the average of the positive and negative field sweeps. The data is smoothly interpolated to avoid the situation where we simulate a field value which is not precisely the same as an experimental field value, and take the root mean square error (RMSE) between the two. This process is then repeated over a large grid of \((D_\mathrm{s}, \tau_\mathrm{s})\) pairs in order to find a global minimum in the difference between simulation and experiment. This optimized fitting is shown in Fig. \ref{fig:HanlePristineSimFit}.

The fitting for the spin transport parameters is comparable to fitting the signal analytically using Eq. \ref{eq:spin_diffusion_equation}, in that the basic principle involves changing $D_\mathrm{s}$, $\tau_\mathrm{s}$, and $P_\mathrm{i}$ until the simulated curve matches the experiment as closely as possible. However, here we fit using just $D_\mathrm{s}$ and $\tau_\mathrm{s}$, which control the shape of the curve, by manually adding the scaling condition that the experimental and simulated values of \(\Delta R_{\mathrm{NL}}\) at \(H_\mathrm{x} = 0\) are the same. This ignores the magnitude of the signal, corresponding to $P_\mathrm{i}$, the interface spin polarization in Eq. \eqref{eq:spin_diffusion_equation}. This method is used for computational convenience. Since $D_\mathrm{s}$ and $\tau_\mathrm{s}$ only affect the shape of the curve, we can obtain them unambiguously while ignoring the magnitude of the signal. This approach allows us to decouple the spin transport parameters from the interface spin polarization, $P_\mathrm{i}$. While this is not exact, it provides a very good approximation and simplifies the analysis significantly, since it allows us to optimize over a 2D ($D_\mathrm{s}, \tau_\mathrm{s}$) search space rather than 3D ($D_\mathrm{s}, \tau_\mathrm{s}$, $P_\mathrm{i}$). To produce the fitted curve in Fig. \ref{fig:HanlePristineSimFit}, we first find $D_\mathrm{s}$ and $\tau_\mathrm{s}$. Once the optimal values are obtained, we fix these in the simulation and remove the normalization condition. Then, using $P_\mathrm{i}$ as the only free parameter, we fit the experimental values. In this case, it is sufficient to use \(\Delta R_{\mathrm{NL}}(H_\mathrm{x} = 0)\) to avoid simulating the entire curve.

\begin{figure}[htpb]
    \centering
    \includegraphics[width=1\linewidth]{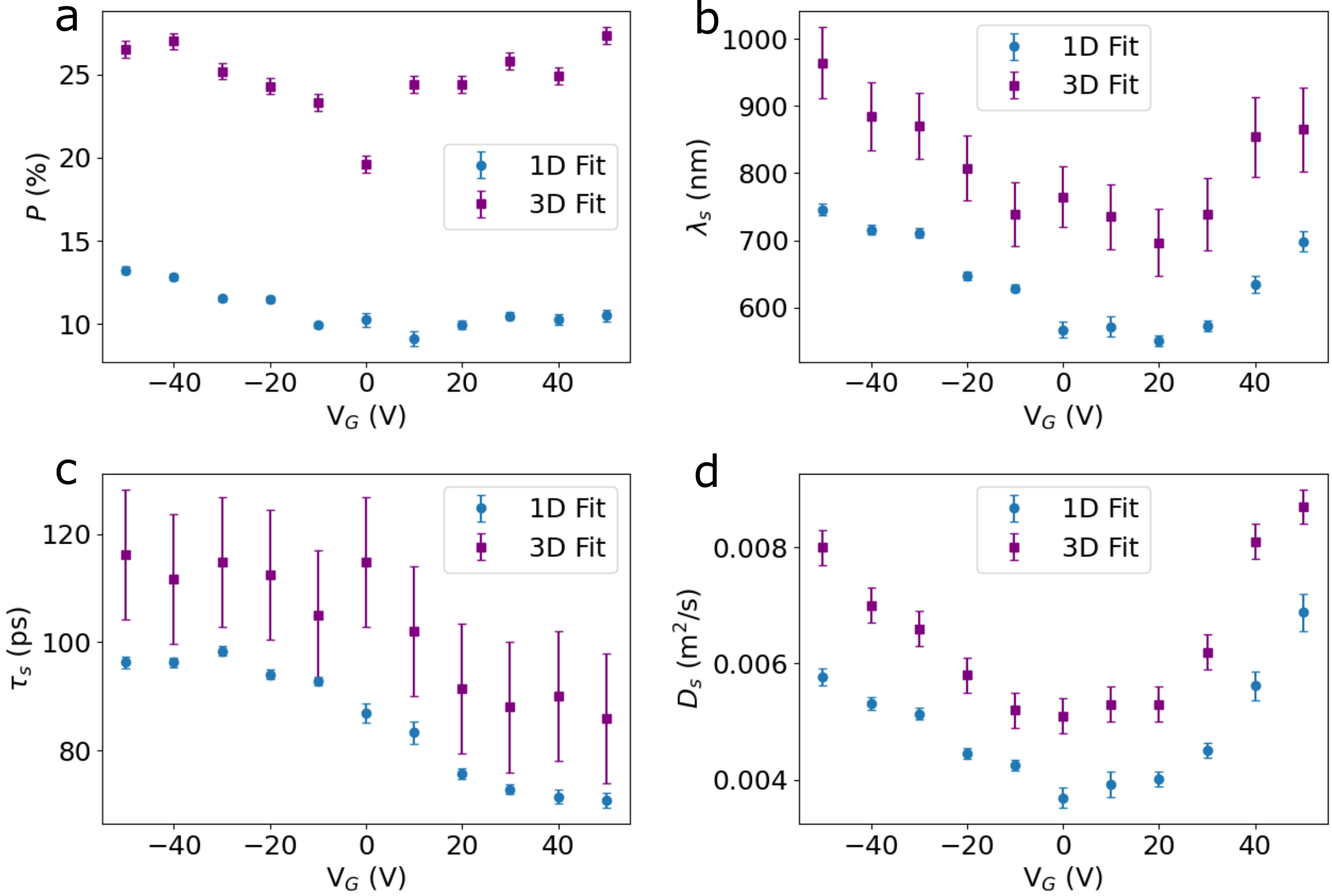}
    \caption{Comparison of the obtained spin transport parameters in the pristine Gr channel via the 1D analytical approach (Eq.~\ref{eq:DeltaRNL}), and the 3D fitting. }
    \label{fig:GrFit1D3D}
\end{figure}

In Fig. \ref{fig:GrFit1D3D}, we summarize the parameters extracted in the Gr region and compare them to those found using the 1D analytical fitting with Eq. \eqref{eq:DeltaRNL} applied to the same data. The results presented here validate the 3D model effectively. First, it is important to note that the two models are not expected to match precisely. This discrepancy arises due to deviations in the device geometry from perfect 1D behavior and the influence of the FM elements on spin accumulation in the channel, which is not considered in the 1D model. Since the FM electrodes provide a pathway for spin relaxation near the contacts, despite the presence of the oxide barrier, we expect a slightly longer \(\lambda_\mathrm{s}\) in the graphene to compensate. This expectation aligns with our observations. 

\begin{figure}[htpb]
    \centering
    \includegraphics[width=1\linewidth]{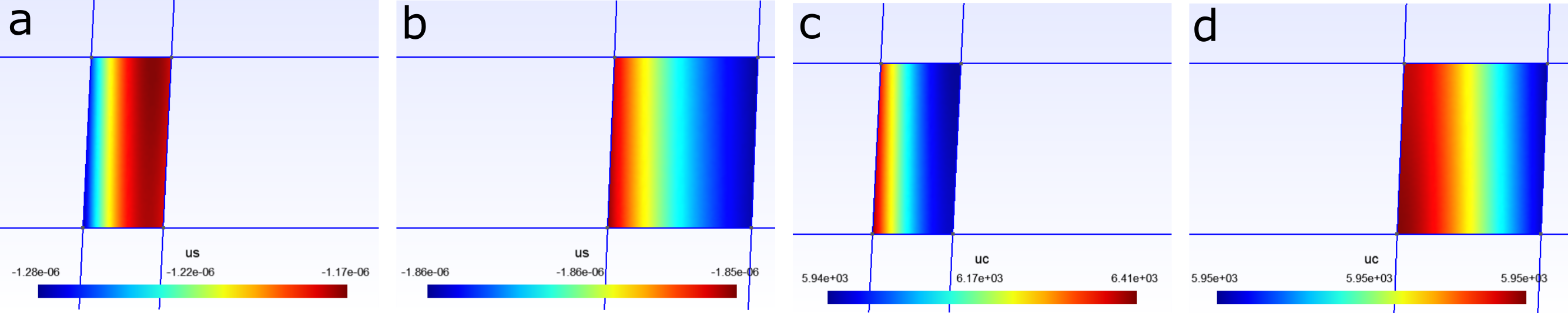}
    \caption{Spin (\(\mu_s\)) (a,b) and charge (\(\mu_c\)) (c,d) chemical potentials as defined by Eqs.~\ref{eq:spin_potentials} and \ref{eq:charge_potential} at the injecting (a,c) and detecting (b,d) faces of the Gr measurement. The values are displayed in arbitrary units, showing the gradient of potentials.}
    \label{fig:InterfaceRainbow}
\end{figure}

The increase in \(\lambda_\mathrm{s}\) is driven by a moderate increase in both $D_\mathrm{s}$ and $\tau_\mathrm{s}$. The interface polarization between the two models is quite different, however, since in the 3D model this corresponds to the value over a 2D region, with a distribution of spin and charge currents, this discrepancy is not unreasonable. Figure~\ref{fig:InterfaceRainbow} shows the gradient of spin and charge potentials at the injecting and detecting interfaces.

 The trend in all cases matches between the two models, even though the absolute values do not. The uncertainty in the 3D model is significantly larger, which is not an inherent property of the approach, but rather a question of practicality, since a finite grid of $D_\mathrm{s}$ and $\tau_\mathrm{s}$values is measured, and the fineness of this search space is limited by computation time. Furthermore, we are not simulating all values of $H_\mathrm{x}$, but rather a discrete selection. Using a longer computation time (Fig. \ref{fig:Error_Colours}), more values of $H_\mathrm{x}$, or indeed a more efficient algorithm could reduce this uncertainty significantly. The current algorithm involves simulating a large square area in the parameter space, but in future works it would be much more sensible to implement a gradient descent approach limiting the number of unhelpful simulations performed, and allowing for a greater degree of automation. 
 \FloatBarrier
 \clearpage
\section{Spin Precession in the Proximitized Graphene}

\begin{figure}[htpb]
    \centering
    \includegraphics[width=0.9\linewidth]{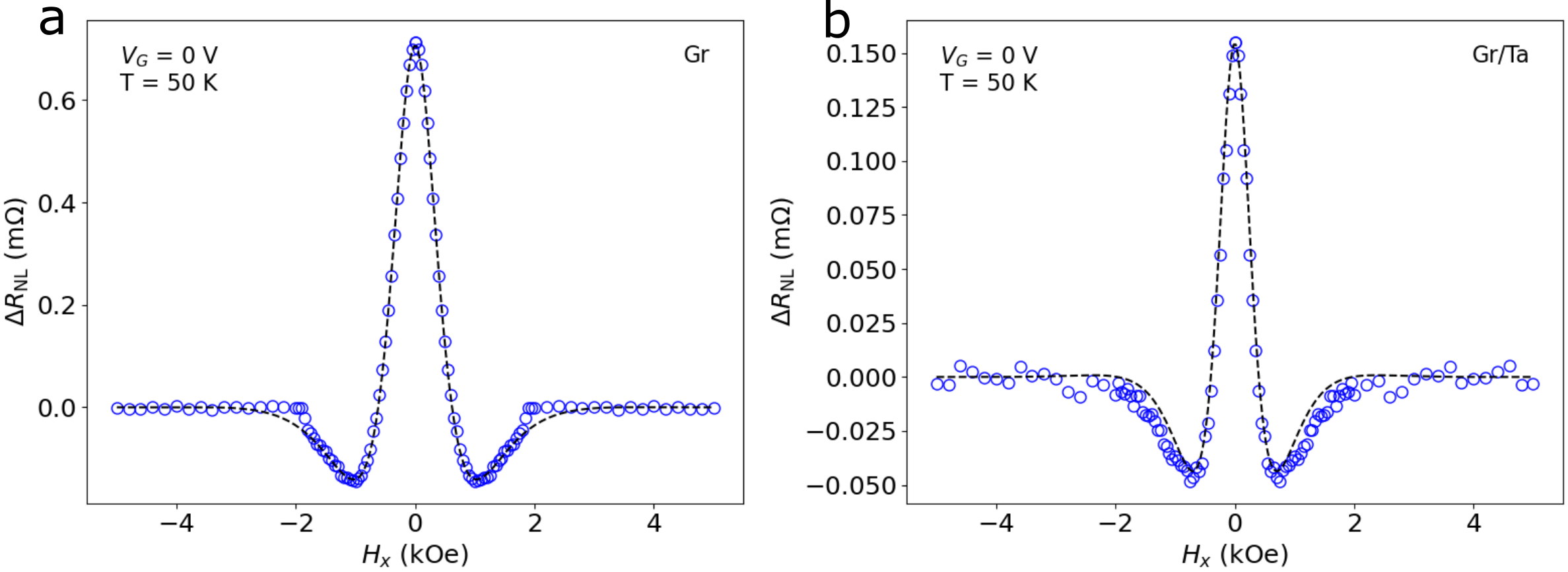}
    \caption{Representative 1D fitting for the Gr (a) and Gr/Ta (b) Hanle measurements. The fit is much better in the case of Gr than Gr/Ta, corresponding to the larger uncertainties in the latter case. All data at 50 K, $V_G=0$}
    \label{fig:1DFitPristProx}
\end{figure}

As in the pristine graphene case, we can fit the spin transport data across the Gr/Ta region using Eq. \ref{eq:DeltaRNL}. This is not an ideal method for the Gr/Ta part in particular, because it assumes a homogeneous 1D channel, thus not accounting for the different regions or the arms of the Hall bar. We compare the fitting of the Gr and Gr/Ta parts using the 1D analytical approach, based on Eq. \ref{eq:DeltaRNL}.  The value for $R_\mathrm{sq}$ used in the Gr/Ta fitting is the effective value over the whole region, and $W$ is the same in both cases—ignoring any effect from the arms of the cross. The idea here is to observe any general trend in the data—for example, if the effective value of \(\lambda_\mathrm{s}\) in the whole crossing region decreases, then we could expect this to be driven by a very large change in the Gr/Ta region. A reference fit using Eq. \ref{eq:DeltaRNL} in each data set is shown in Fig. \ref{fig:1DFitPristProx}, to illustrate the reason for the larger uncertainty in the Gr/Ta case.

\begin{figure}[htpb]
    \centering
    \includegraphics[width=1\linewidth]{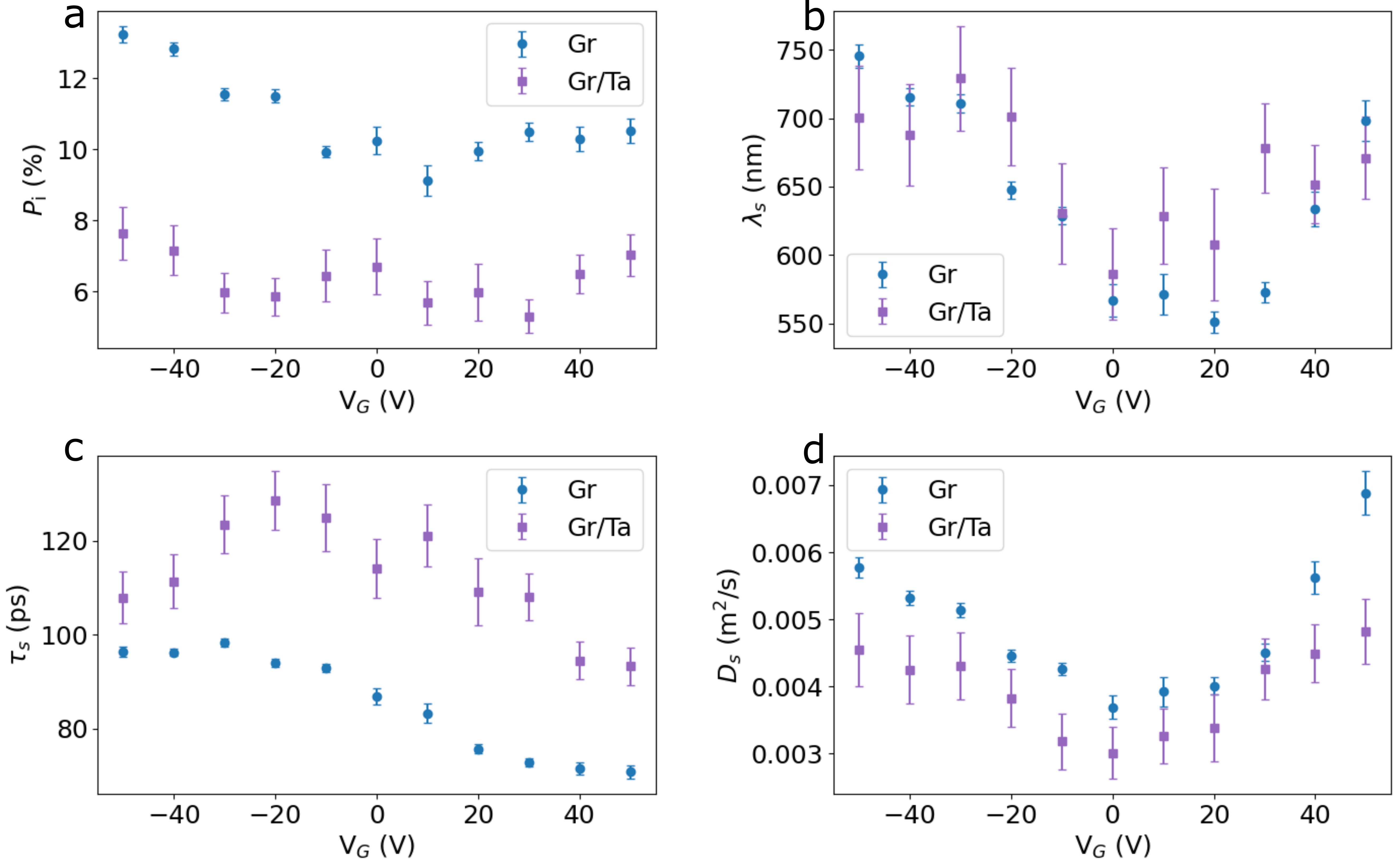}
    \caption{Spin transport parameters fitted from the 1D analysis of the Hanle curves in the Gr and Gr/Ta regions, using Eq.~\ref{eq:DeltaRNL}, with the appropriate values of $R_\mathrm{sq}$, $L$, and $W$.}
    \label{fig:1DFitPristProxResults}
\end{figure}

A summary of the spin transport parameters extracted from the Gr and Gr/Ta measurements via fitting to Eq.~\ref{eq:DeltaRNL} is shown in Fig.~\ref{fig:1DFitPristProxResults}. Firstly, $P_\mathrm{i}$ is significantly different. At \( V_{G} = 0 \, \text{V} \), we have \( 10.2 \pm 0.4 \, \% \) for the Gr and \( 6.7 \pm 0.7 \, \% \) for the Gr/Ta. This difference can be primarily attributed to the use of different contacts for each measurement, and since each interface behaves differently, substantial changes to the interface polarization can arise, even for contacts deposited in the same device. This difference in interface nature can be readily observed from the varying interface resistance values, with the three measured interfaces having resistances of \( 9.0 \, \text{k}\Omega \), \( 1.4 \, \text{k}\Omega \), and \( 22.5 \, \text{k}\Omega \) for F1, F2, and F3, respectively (Fig.~\ref{fig:SEM}). Although there is no clear relationship between $P_\mathrm{i}$ and the interface resistance, such large resistance differences imply a significant variation in either the quality or thickness of the TiO\(_x\) layer, which can be expected to affect $P_\mathrm{i}$. This is a key source of uncertainty in our estimate of \( \theta_{\mathrm{SH}} \), since both $P_\mathrm{i}$ and \( \theta_{\mathrm{SH}} \) have the same effect on the signal (Eq. \ref{eq:DeltaRSCI}), making it impossible to disentangle the two. This is evident in Eq.~\ref{eq:DeltaRSCI}, which provides an analytical description of SCI measurements, albeit with similar limitations to Eq.~\ref{eq:DeltaRNL} for spin transport.

\begin{figure}[htpb]
    \centering
    \includegraphics[width=1\linewidth]{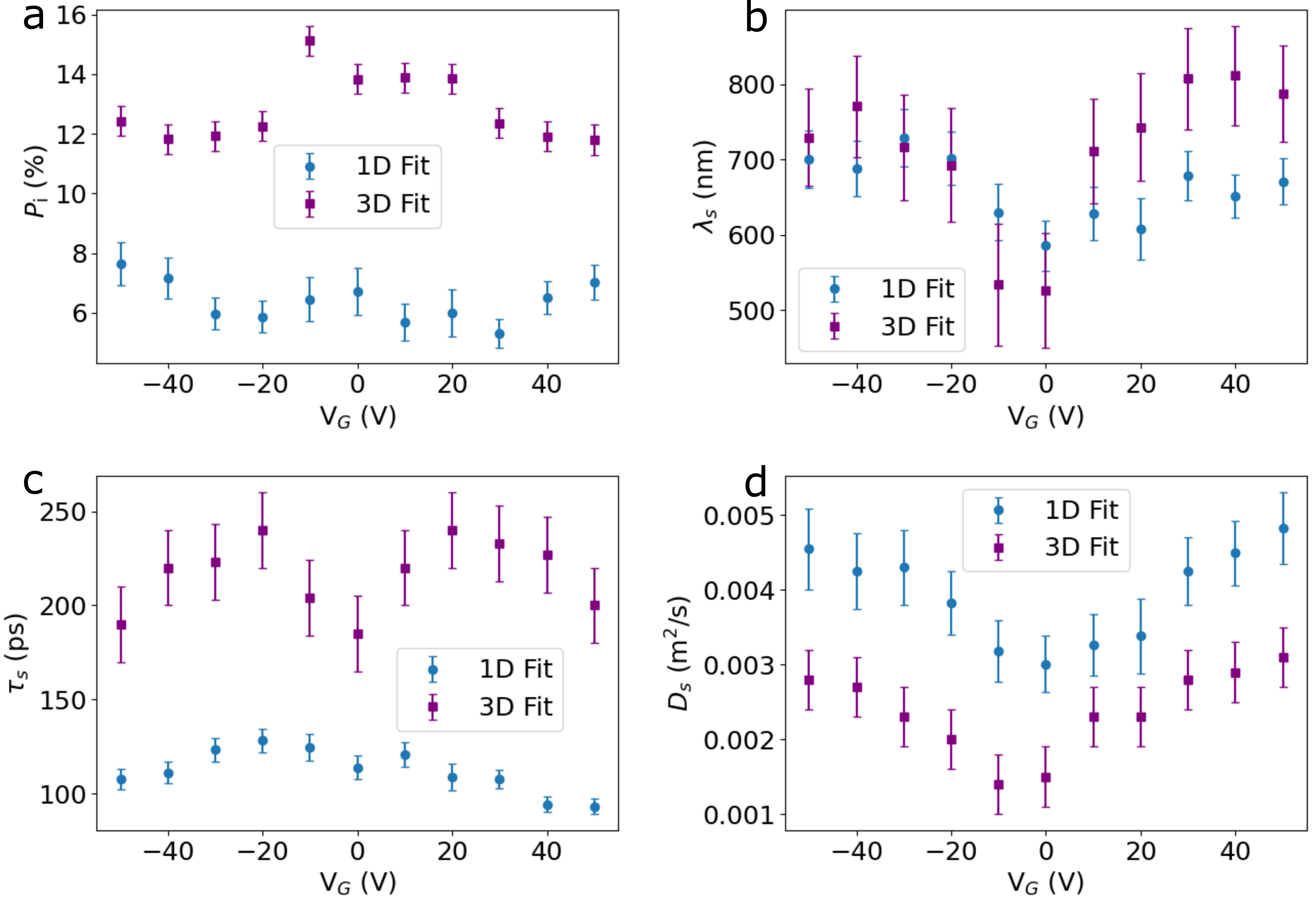}
    \caption{Summary of the spin transport parameters obtained from 1D and 3D fitting of the Gr/Ta region.}
    \label{fig:SI9}
\end{figure}

In the spin transport measurements, the signal also depends equally on $P_\mathrm{i}$ from the injector and detector. Therefore, for analysis, we must assume they are the same. However, it is evident even from this simple fitting that this is not the case. For the SCI measurements, we use one electrode from the Gr/Ta spin transport data, which limits the uncertainty, but in principle, $P_\mathrm{i}$ for the injector and detector could be quite different. Fig. \ref{fig:1DFitPristProxResults}b shows that \(\lambda_\mathrm{s}\) for both regions is quite similar, albeit with larger uncertainty for the Gr/Ta. This is due to a reduced signal to noise ratio, as shown in the two fits in Fig. \ref{fig:1DFitPristProx}, and applies to all extracted parameters using both 1D and 3D approaches. A similar value of \(\lambda_\mathrm{s}\) is reasonable in this case, as the region being measured consists primarily of pristine graphene (Fig. \ref{fig:SEM}b). The parameters in the Gr/Ta region would need to be significantly modified to produce visible differences in the effective parameters, so from this simplified 1D model we would not expect a large difference in $\lambda_\mathrm{s}$ in the isolated Gr/Ta region.

\(\lambda_\mathrm{s}\) is typically the parameter of primary interest, but it is defined in terms of $D_\mathrm{s}$ and $\tau_\mathrm{s}$. When examining $D_\mathrm{s}$ and $\tau_\mathrm{s}$ individually, there is a clear difference. $\tau_\mathrm{s}$ is larger in the Gr/Ta data (\(87 \pm 2 \, \text{ps}\) for Gr vs. \(114 \pm 6 \, \text{ps}\) for Gr/Ta at \(V_G = 0 \, \text{V}\)). This difference is compensated by the difference in $D_\mathrm{s}$, which is found to be smaller in the Gr/Ta measurement \((3.69 \pm 0.18) \times 10^{-3} \, \text{m}^2/\text{s}\) vs. \((3.01 \pm 0.38) \times 10^{-3} \, \text{m}^2/\text{s}\) at \(V_G = 0 \, \text{V}\), resulting in a similar value for \(\lambda_\mathrm{s}\) over the entire range of \(V_G\). Next, we apply the 3D simulation, which allows us to separate the two regions precisely and determine values for the Gr/Ta part of the device, rather than just effective parameters. To find \(D_\mathrm{s}^{\mathrm{Gr/Ta}}\) and \(\tau_\mathrm{s}^{\mathrm{Gr/Ta}}\) independently, we fix the values of \(D_\mathrm{s}^{\mathrm{Gr}}\) and \(\tau_\mathrm{s}^{\mathrm{Gr}}\) in the pristine part of the Gr/Ta measured area within the 3D model, taking these parameters from the 3D simulation of the Gr data in Fig. \ref{fig:GrFit1D3D}. We then define boundary conditions corresponding to the applied current and measure the simulated voltage, following the procedure described previously. Finally, we scan a range of \(D_\mathrm{s}^{\mathrm{Gr/Ta}}\) and \(\tau_\mathrm{s}^{\mathrm{Gr/Ta}}\) until we achieve the best fit to the experimental data from the simulation. The obtained values are summarized in Fig. \ref{fig:SI9} and compared to the effective values obtained from the 1D model in the Gr/Ta region.

\begin{figure}[htpb]
    \centering
    \includegraphics[width=1\linewidth]{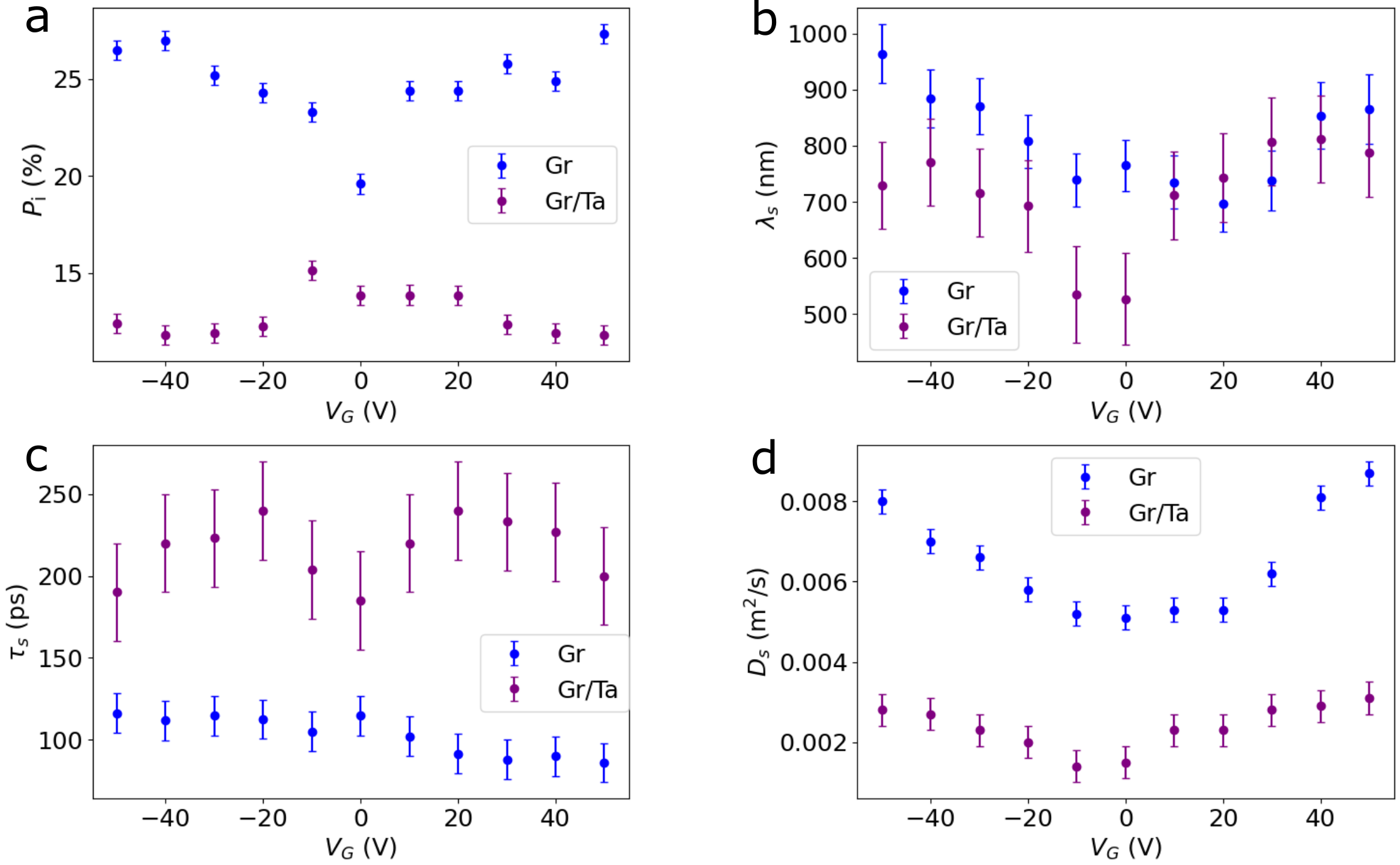}
    \caption{Summary of spin transport parameters obtained from 3D fitting of the Gr and Gr/Ta region.}
    \label{fig:3D_Values_All_Summary}
\end{figure}

Finally, we present the full 3D simulation comparing both regions, Gr and Gr/Ta (Fig. \ref{fig:3D_Values_All_Summary}). Results show that there is a general reduction in \(\lambda_\mathrm{s}\) in the Gr/Ta region, but the reduction is rather small, and indeed at positive gate voltages the two begin to overlap. The origin of this reduction is mixed — we have opposite contributions from $D_\mathrm{s}$ and $\tau_\mathrm{s}$, with $\tau_\mathrm{s}$ actually increasing in the Gr/Ta region compared to the pristine, but this is offset by a much larger change of $D_\mathrm{s}$ in the opposite direction, resulting in an overall decrease in \(\lambda_\mathrm{s}\). This is in keeping with the results from the 1D model in Fig. \ref{fig:1DFitPristProxResults}.

Explaining the exact physical origin of these effects is difficult. $D_\mathrm{s}$ is closely connected to the conductivity of the graphene, since it dictates the rate of diffusion of electrons. One possible interpretation here is that, as $D_\mathrm{s}$ decreases, the rate of electron-defect interactions is reduced, thus giving a longer $\tau_\mathrm{s}$. Under the Dyakonov-Perel mechanism, the spin relaxation time $\tau_\mathrm{s}$ is inversely proportional to the momentum scattering time $\tau_\mathrm{p}$, i.e., $\tau_\mathrm{s} \propto 1/\tau_\mathrm{p}$ \cite{dyakonov1972spinxxx, Dzhioev2004-toxxx}. Since $D_\mathrm{s} \propto \tau_\mathrm{p}$, a decrease in $D_\mathrm{s}$ corresponds to a reduction in $\tau_\mathrm{p}$, which leads to an increase in $\tau_\mathrm{s}$ \cite{zutic_spintronics_2004xxx} in the Dyakonov-Perel regime, where frequent momentum scattering interrupts the precession of spins around internal effective magnetic fields, thereby reducing the overall dephasing.

The conductivity in the Gr/Ta region is found to increase significantly compared to the Gr (Fig.~\ref{fig:rhosimresults}), despite $D_\mathrm{s}$ being reduced. This indicates a doping effect due to the TaO\(_x\), since the conductivity goes up even though the electron mobility goes down. It is possible that damage to the Gr/Ta region during sputtering causes this reduced mobility, but this is compensated by increased doping.

\FloatBarrier
\clearpage
\section{Spin-charge interconversion in the proximitised graphene}
\begin{figure}[htpb]
    \centering
    \includegraphics[width=0.7\linewidth]{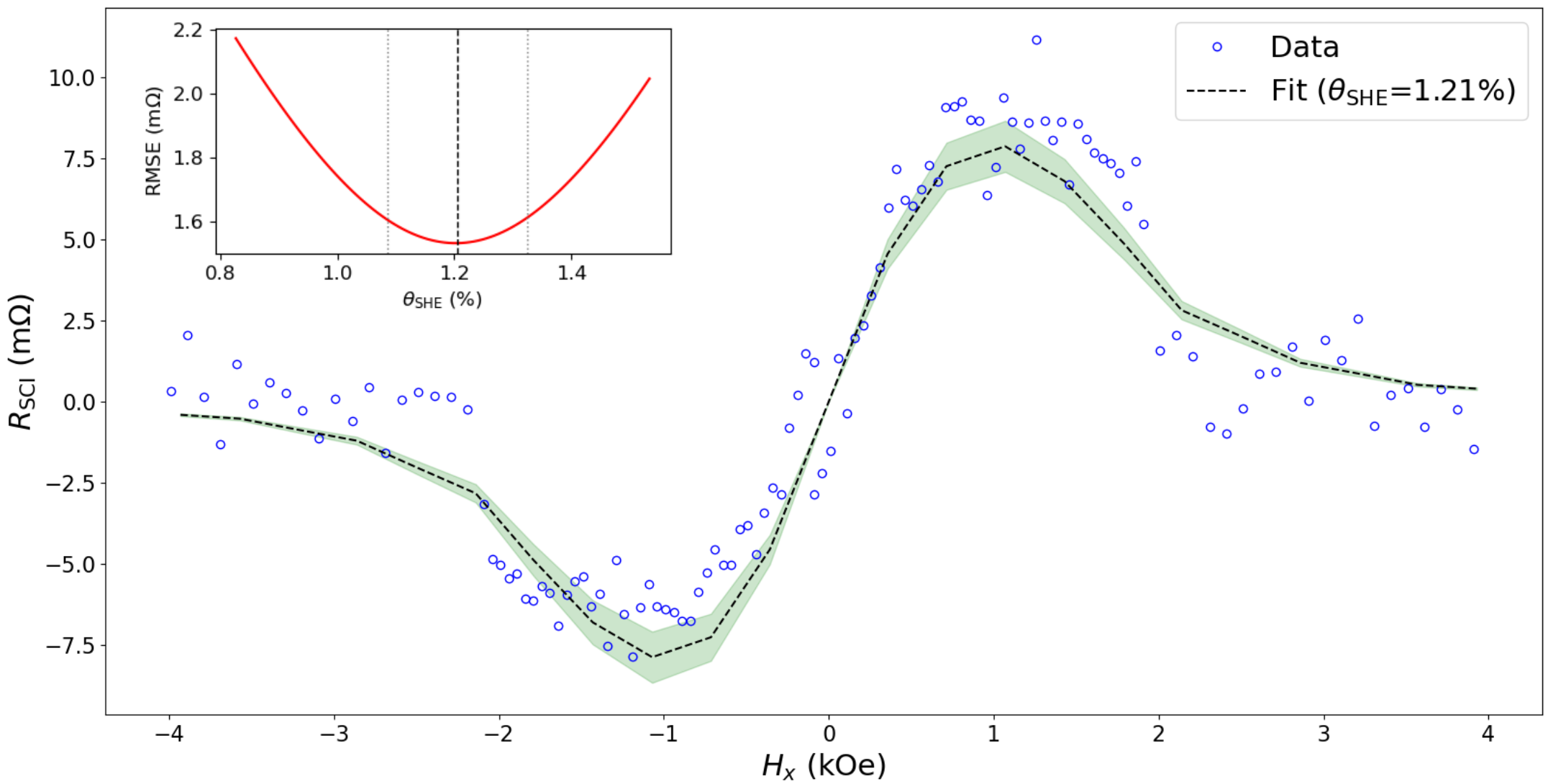}
    \caption{Representative fit of the SCI data in Gr/Ta using the 3D FEM model. Experimental data at $V_G = 0 \, \text{V}$ and $T = 50 \, \text{K}$ is shown, along with the simulated best fit. The inset shows the variation of the RMSE between the fit and experiment as a function of $\theta_\text{SH}$. The confidence interval is taken as the minimum of the RMSE $\pm$ 10\%, corresponding to the green shaded area.}
    \label{fig:SCI}
\end{figure}

Having obtained \(D_\mathrm{s}^{\mathrm{Gr}}\), \(\tau_\mathrm{s}^{\mathrm{Gr}}\), \(D_\mathrm{s}^{\mathrm{Gr/Ta}}\), and \(\tau_\mathrm{s}^{\mathrm{Gr/Ta}}\), as well as the corresponding $P_\mathrm{i}$ values, from the Hanle measurement, we now move on to the final step of obtaining \(\theta_{\mathrm{SH}}\). This is done using the SCI data at 50\,K and $V_\mathrm{G}$. In principle, if our model is consistent, we should be able to fit the SCI data using only \(\theta_{\mathrm{SH}}\) as a free parameter. We perform the 3D simulation, using the off-diagonal conductivity in the Gr/Ta region (Eq. \eqref{eq:she_y_up}) defined by \(\theta_{\mathrm{SH}}\) as the only fitting parameter. This fitting is shown in Fig. \ref{fig:SCI} for the 50 K $V_G=0$ SCI data. The error shown in green corresponds to the $\theta_\mathrm{SH}$ at the minimum of the RMSE between the 3D model and the interpolated experimental data \(\pm 10 \%\) (See Note \ref{Uncertainty Analysis}). The value for $P_\mathrm{i}$ used here is that derived from the spin transport data in Gr/Ta fitted to the 3D model (Fig. \ref{fig:3D_Values_All_Summary}).

\begin{figure}[htpb]
    \centering
    \includegraphics[width=1\linewidth]{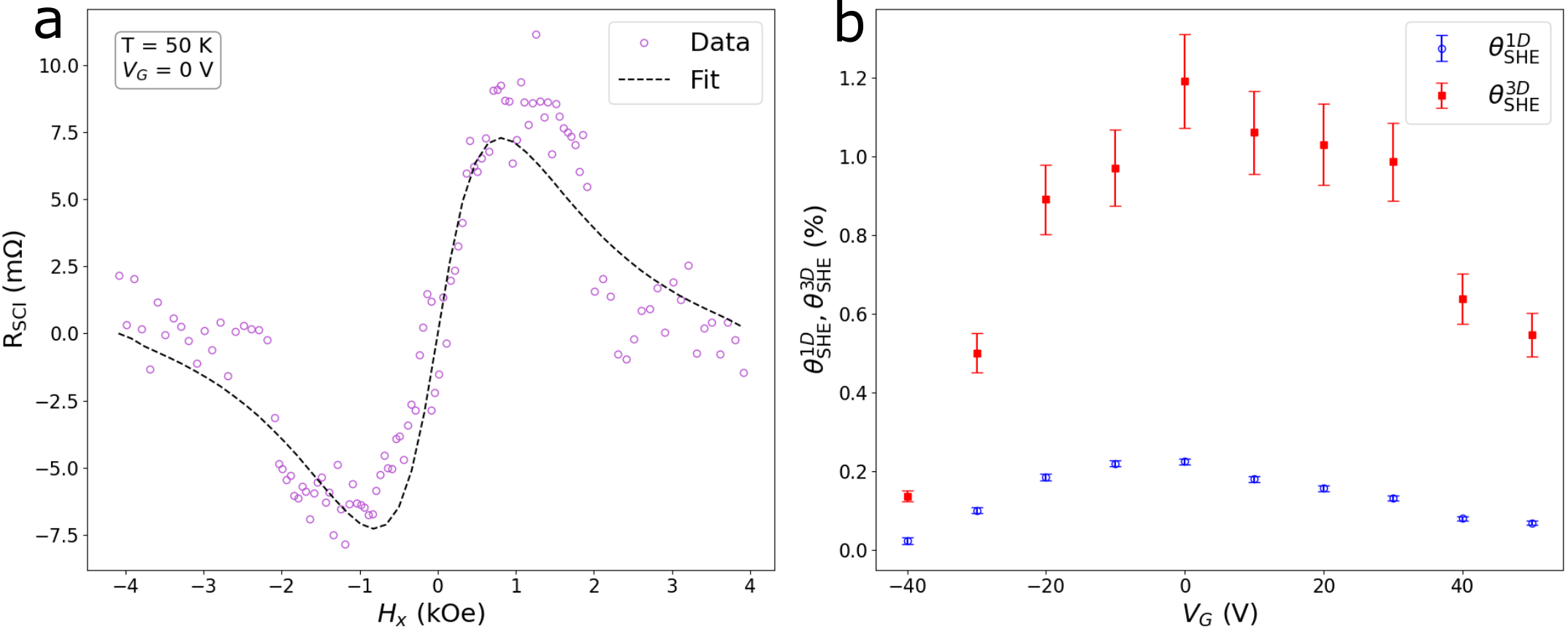}
    \caption{(a) Representative fit of the SCI data to the 1D analytical model from Eq.~\eqref{eq:DeltaRSCI}, with $\theta_\text{SH}$ as the fitting parameter and using values for $D_\mathrm{s}$, $\tau_\mathrm{s}$, and $P_\mathrm{i}$ obtained from the 1D fit of the spin transport data in the Gr/Ta region. (b) Comparison of the resultant values for $\theta_\text{SH}$ from the 1D and 3D models, demonstrating a similar trend but a systematically much larger value obtained via the 3D model. All data at 50 K.}
    \label{fig:SCI_Compare1D3D}
\end{figure}

The alternative approach, without using 3D simulation, is to fit the data using the analytical 1D model from Eq.~\eqref{eq:DeltaRSCI}. This can be done either by fixing all parameters from the spin diffusion measurements and fitting the data with $\theta_\text{SH}$ or by using $\tau_\mathrm{s}$and $D_\mathrm{s}$ in addition to $\theta_\text{SH}$ for fitting the data. We can achieve a good fit using just $\theta_\text{SH}$, both in 1D and 3D, so this is the approach we use to compare the models. For the 1D fitting, the parameters involved are 'effective' values consisting of a mix of Gr and Gr/Ta regions. A representative fit of the SCI data is shown in Fig.~\ref{fig:SCI_Compare1D3D}a, where the 1D model does fit the data well. However, the parameters returned differ significantly from those found in the 3D model (Fig. \ref{fig:SCI_Compare1D3D}), which considers distinct spin transport parameters and an accurate device geometry.

\FloatBarrier
\clearpage
\section{Uncertainty Analysis in 3D Model}
\label{Uncertainty Analysis}
\begin{figure}[htpb]
    \centering
    \includegraphics[width=1\linewidth]{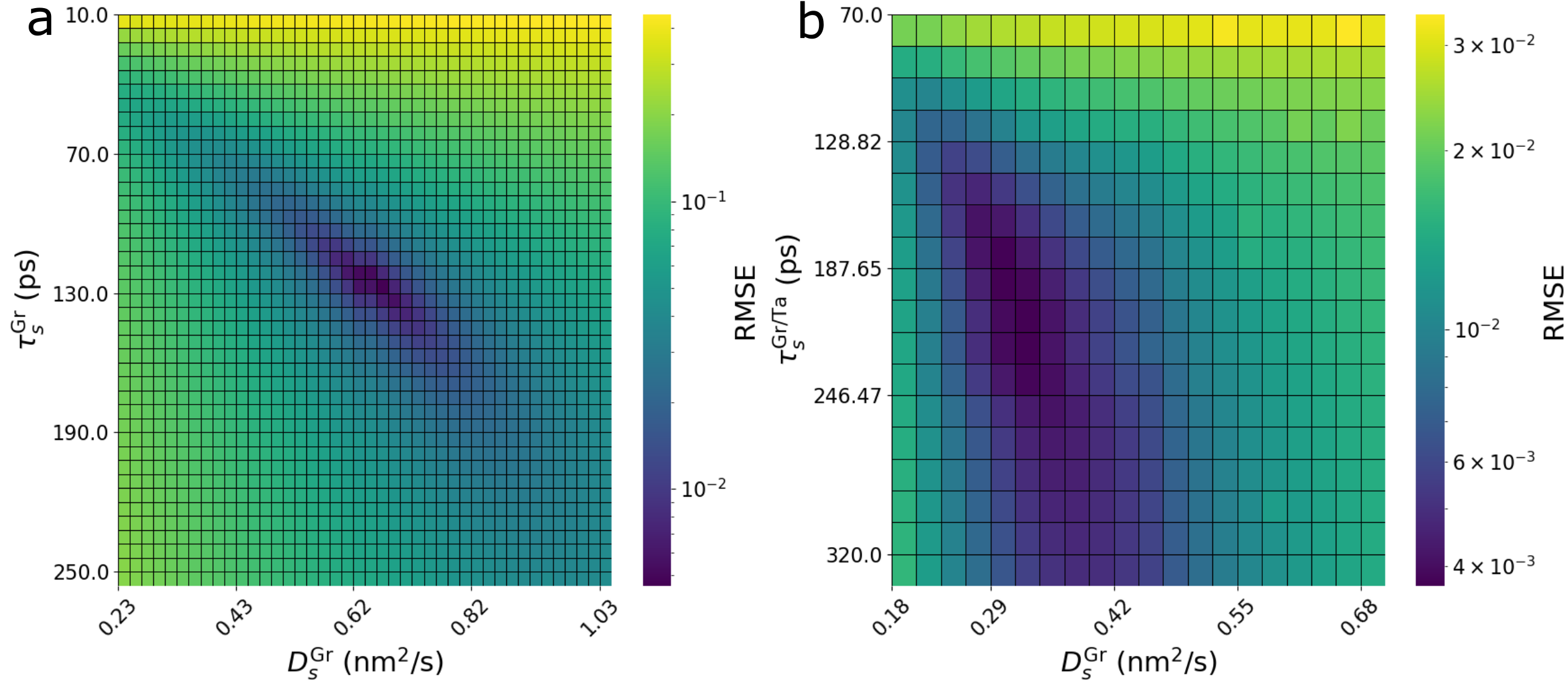}
\caption{
a) Scan of the \(D_\mathrm{s}^{\text{Gr}}\) and \(\tau_\mathrm{s}^{\text{Gr}}\) parameter space in the Gr spin transport data at 50 K and \(V_G = 40 \, \text{V}\), showing the RMSE on a logarithmic scale. The dark spot in the center corresponds to the global minimum and the values shown for the 3D simulation at this \(V_G\). b) The same parameter space scan for \(\tau_\mathrm{s}^{\text{Gr/Ta}}\) and \(D_\mathrm{s}^{\text{Gr/Ta}}\), again showing the region scanned and the global minimum. In this case, the simulations are significantly more computationally intensive, and thus the resolution scanned is lower.}
    \label{fig:Error_Colours}
\end{figure}

In order to explore the accuracy of the 3D model, and ensure that the observed solution was a global minimum, we simulated the spin transport over a large range of $D_\mathrm{s}$ and $\tau_\mathrm{s}$values in the pristine (Fig. \ref{fig:Error_Colours}a) and proximitized (Fig. \ref{fig:Error_Colours}b) systems. The plots show the root mean square error (RMSE) between the experimental data and the simulation over the parameter space explored. The Gr measurement occurs over a smaller, simpler region and therefore is much faster to simulate, allowing us to plot a comparatively large region in high resolution. This clearly shows the global minimum corresponding to the values in Fig. \ref{fig:3D_Values_All_Summary}. This also allows us to quantify the error in the simulation, with the error bars in Fig. \ref{fig:3D_Values_All_Summary} corresponding to the \((D_\mathrm{s}, \tau_\mathrm{s})\) values with the minimum RMSE plus a region around the minimum defined as \(\mathrm{RMSE}_{\text{min}} \pm 0.1 \times \mathrm{RMSE}_{\text{min}}\). For the pristine case, this corresponds to \(\tau_\mathrm{s} \pm 30 \, \text{ps}\) in the pristine region and \(\tau_\mathrm{s} \pm 50 \, \text{ps}\) in the proximitized region. For $D_\mathrm{s}$, the corresponding values are \((D_\mathrm{s} \pm 0.03) \times 10^{16}~\mathrm{nm}^2/\mathrm{s}\) in the Gr region and \((D_\mathrm{s} \pm 0.04) \times 10^{16}~\mathrm{nm}^2/\mathrm{s}\) in the Gr/Ta region. As can be seen in Fig. \ref{fig:1DFitPristProx}, the Gr/Ta data in general is not as clear as the Gr data, with a lower signal-to-noise ratio and a less precise fit using both the 1D and 3D approaches.

The uncertainty in the SCI data fitting is discussed above, but one key source of uncertainty is the value of $P_\mathrm{i}$. We take the value from the spin transport measurement in the Gr/Ta region, however, as seen in both the 1D and 3D fitting, this value varies by a factor of 2 between the Gr and Gr/Ta measurements. The value used to fit the SCI data comes from a measurement using one of the same electrodes used in the SCI measurement, meaning it is the best possible estimate, but nonetheless this could lead to significant variance in the value found in \(\theta_{\mathrm{SH}}\). Another key source of error, which is difficult to estimate, is the assumed homogeneity of the graphene. We have assumed throughout that both the pristine graphene and the Gr/Ta regions have different parameters governing their behaviour, but that these parameters (\(\rho, D_\mathrm{s}, \tau_\mathrm{s}\)) are perfectly homogeneous internally. This is a necessary assumption to perform such an analysis, but it cannot be ruled out that different regions of the graphene, either inherently or due to defects induced during fabrication, possess different spin transport properties.

\FloatBarrier
\clearpage
\section{Reproducibility}
\label{Second_Ta_Device_Reproducibility}

\begin{figure}[htpb]
    \centering
    \includegraphics[width=1\linewidth]{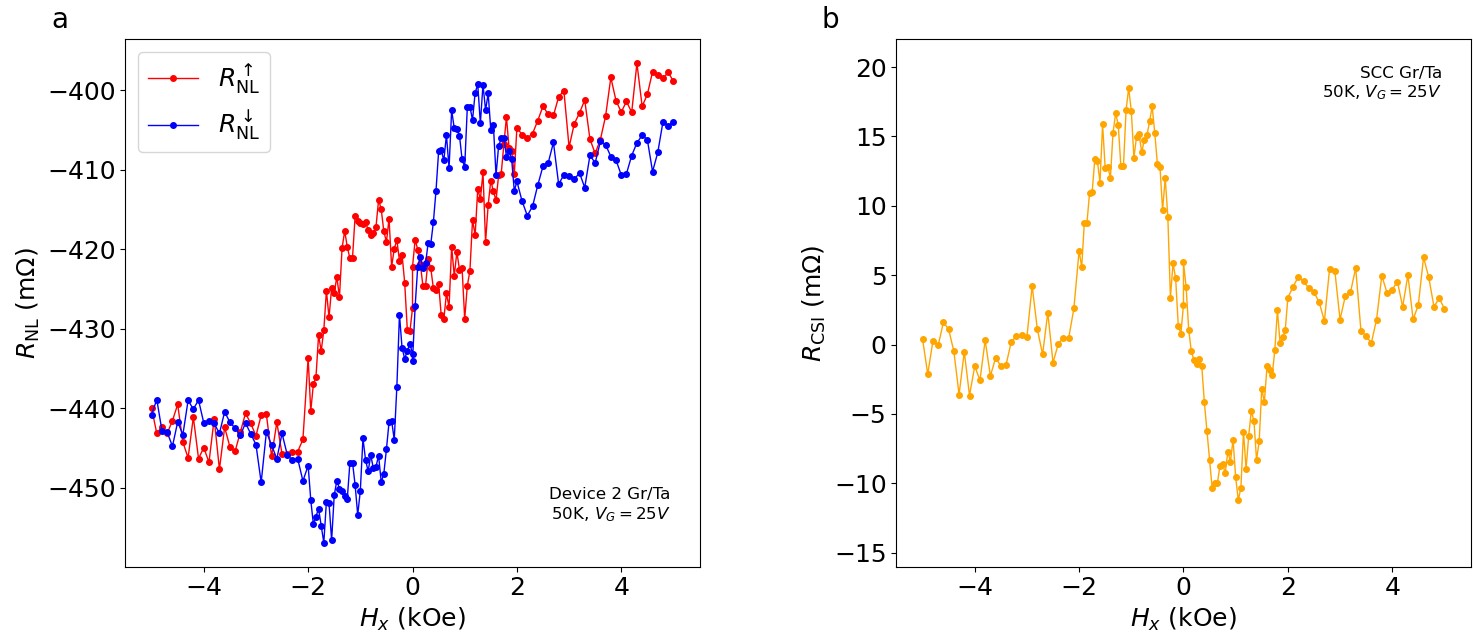}
    \caption{(a) Non-local resistance $R_\mathrm{NL}$ for CSI measured as a function of $H_\mathrm{x}$ in a second Gr/Ta heterostructure at 50~K with an external gate voltage of $V_\mathrm{G}=25 V$. This compares closely with Fig. 4 b,c in the main text. The red (\( R_{\text{NL}}^\uparrow \)) and blue (\( R_{\text{NL}}^\downarrow \)) curves represent measurements with the initial magnetization of the FM electrodes along the \( +y \) and \( -y \) directions, respectively. (b) The CSI resistance \( R_{\text{CSI}} \) (yellow) extracted from the difference of the red and blue curves in (a).}
    \label{fig:Reproduc_Fig}
\end{figure}

In order to demonstrate reproducibility of these results, we fabricated a second Gr/TaO$_x$ device, following the same method. This device gave less complete results and did not permit the characterisation of the pristine graphene in the same device due to broken electrical contacts. However, we were able to clearly measure the SHE, with a similar strength to the device discussed in the main text. The results at 50\,K and $V_\mathrm{G}=25 V$ are shown in Fig. \ref{fig:Reproduc_Fig}.

\FloatBarrier
\clearpage
\section{Measuring CSI with out-of-plane field}
\label{CSI out of plane field}

\begin{figure}[htpb]
    \centering
    \includegraphics[width=0.65\linewidth]{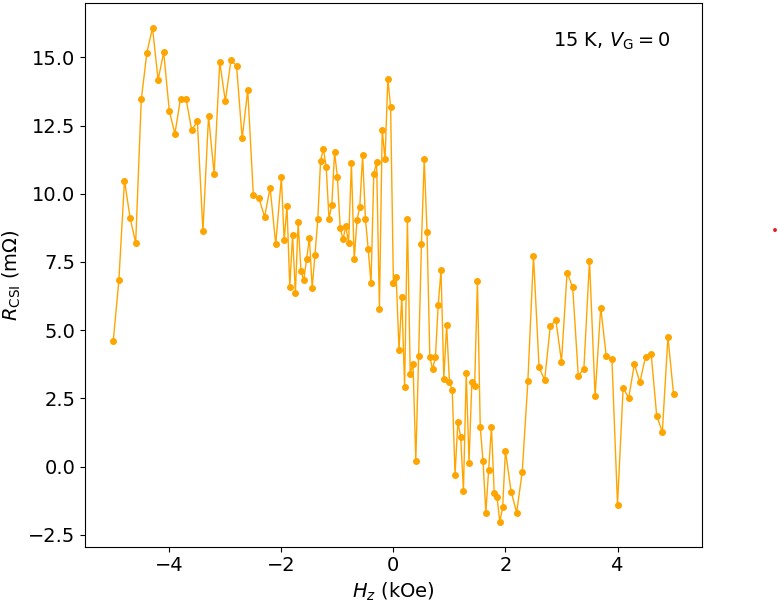}
    \caption{The CSI resistance, \( R_{\text{CSI}} \), extracted from the difference of \( R_{\text{NL}}^\uparrow \) and \( R_{\text{NL}}^\downarrow \) at 15\,K and no applied gate voltage, as a function of $H_z$. The absence of precession (compared to Fig. 4c in the main text for example), indicates there is no REE in this system.}
    \label{fig:Hz}
\end{figure}

To investigate the origin of the S-shaped background in Fig. 4b of the main text, we measure CSI with an out-of-plane ($H_z$) field. If the origin of the background is due to the REE, we would expect to see a precessing CSI signal due to the oscillating x component of the spin current. However, this is not the case. As visible in Fig. \ref{fig:Hz}, we show $R_\mathrm{CSI}$ with no visible signal from precession, which demonstrates an absence of REE, indicating that the background signal is due to charge effects, rather than spin.

\FloatBarrier
\clearpage
\section{Measuring CSI at high temperature}

\begin{figure}[htpb]
    \centering
    \includegraphics[width=0.65\linewidth]{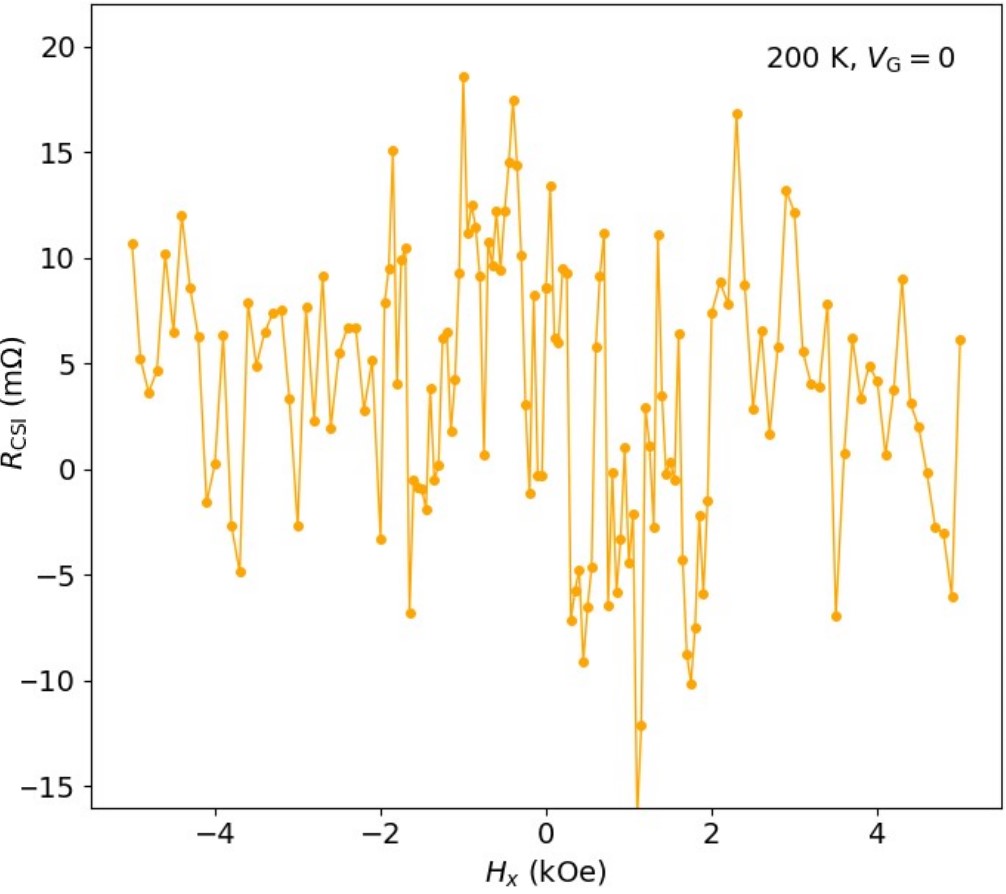}
    \caption{The CSI resistance, \( R_{\text{CSI}} \), extracted from the difference of \( R_{\text{NL}}^\uparrow \) and \( R_{\text{NL}}^\downarrow \) at 200\,K and no applied gate voltage, with no visible precession signal.}
    \label{fig:200K}
\end{figure}

As stated in the main text, we were unable to measure SCI above 150 K. Figure \ref{fig:200K} shows the experimental data for the same CSI measurement as shown in Fig. 4c of the main text, but at 200\,K, showing no evidence of CSI, although a signal might be present and unresolvable due to the greater thermal noise.

\FloatBarrier
\clearpage
\section{Comparing CSC and SCC}
\label{CSI 200K}

\begin{figure}[htpb]
    \centering
    \includegraphics[width=1\linewidth]{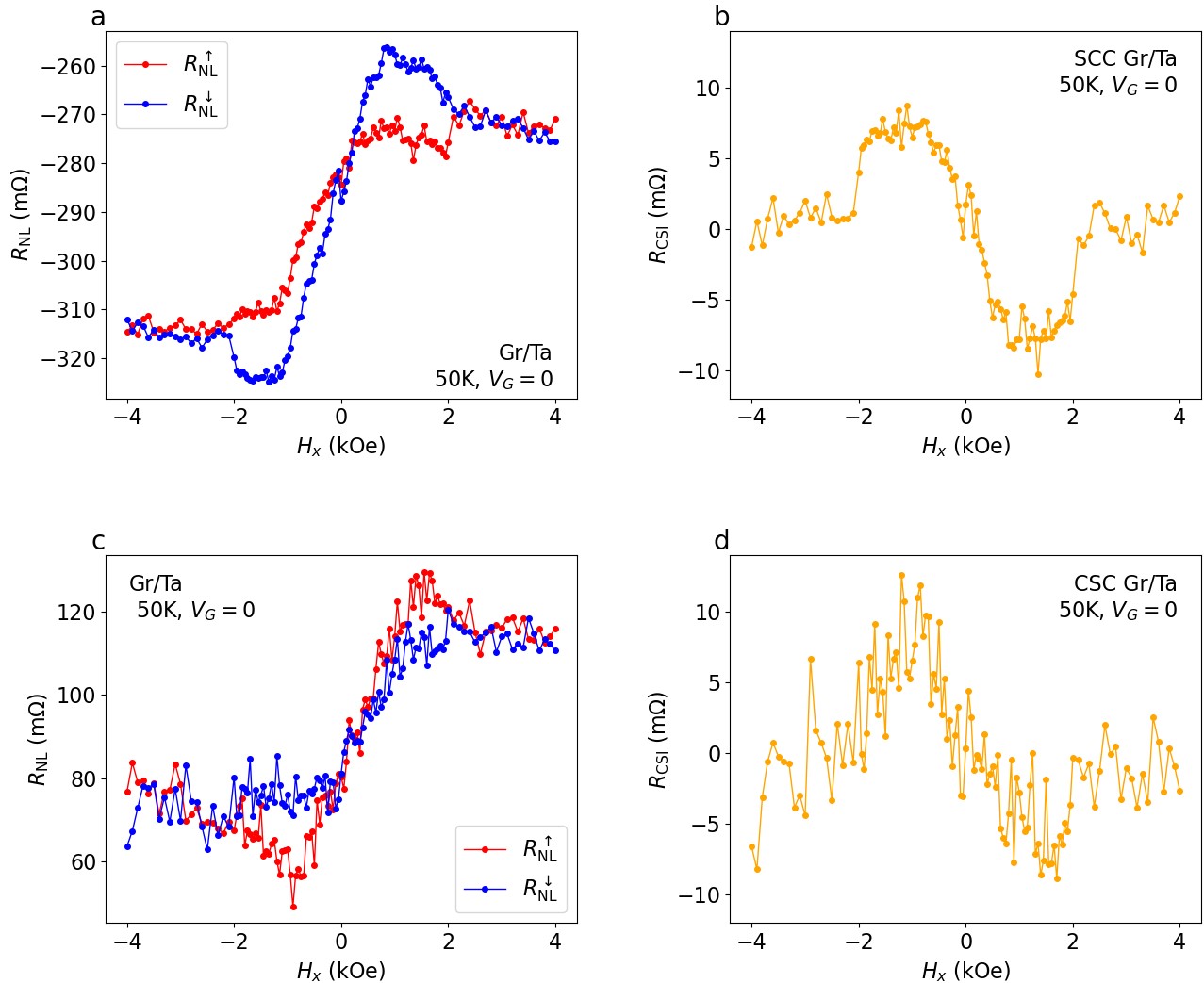}
    \caption{\( R_{\text{NL}}^\uparrow \) and \( R_{\text{NL}}^\downarrow \) in (a) the SCC measurement configuration and (c) the CSC configuration, with the corresponding \( R_{\text{CSI}} \) for (b) SCC and (d) CSC. All data are at 50\,K and zero applied gate voltage.}
    \label{fig:CSC_SCC}
\end{figure}

It is possible to measure CSI in two different configurations, either spin to charge conversion (SCC) as used in the main text (Fig.\,4\,b,c), or charge to spin conversion (CSC). The two are linked by Onsager reciprocity, and experimentally the difference is in switching the current and voltage probes. A comparison is shown in Fig. \ref{fig:CSC_SCC}, where the results are similar but the SCC measurement has lower noise, which is why this configuration is used in the main text.

\end{document}